\documentclass[sn-standardnature,iicol]{sn-jnl}
\usepackage{cuted}[noshrink]
\jyear{2022}%
\theoremstyle{thmstyleone}%
\theoremstyle{thmstyletwo}%

\theoremstyle{thmstylethree}%

\usepackage{comment}
\usepackage{lineno}
\usepackage{lipsum}
\usepackage{textcomp}
\usepackage{soul}
\makeatletter
\newcommand{\ssymbol}[1]{{\@fnsymbol{#1}}}
\makeatother

\usepackage{comment,mfirstuc}
\usepackage{xspace}
\usepackage{tabularx}
\usepackage{array} 
\usepackage{booktabs} 

\usepackage{adjustbox}
\usepackage{longtable}
\usepackage{hyperref}

\definecolor{aureolin}{rgb}{0.99, 0.93, 0.0}
\definecolor{amber}{rgb}{1.0, 0.49, 0.0}
\newcolumntype{C}[1]{>{\centering\small\arraybackslash}p{#1}}
\newcolumntype{P}[1]{>{\centering\small\arraybackslash}p{#1}}

\newcommand{\epic}{{ePIC}\xspace}

\newcommand{\gluex}{{\textsc{GlueX}}\xspace}

\newcommand{\addComment}[2]{
  \expandafter\newcommand\csname #1\endcsname[1]{{\bf \color{#2} \capitalisewords{#1}:\,##1}}
  \expandafter\newcommand\csname #1cor\endcsname[2]{{\color{#2} \capitalisewords{#1}:\,\st{##1}{\bf ##2}}}
  \expandafter\newcommand\csname #1color\endcsname{#2}
}

\addComment{cris}{blue} 
\addComment{evaristo}{red}
\addComment{tanja}{magenta}
\addComment{tmp}{blue} 
\addComment{james}{green}
\addComment{karthik}{amber}

\DeclareUnicodeCharacter{2212}{-}

\begin{document}

\title[AI4EIC]{Artificial Intelligence for the Electron Ion Collider (AI4EIC)}

\author[61]{\fnm{C.} \sur{Allaire}}\equalspeak{Invited speaker}
\author[24]{\fnm{R.} \sur{Ammendola}}\equalspeak{Invited speaker}
\author[3]{\fnm{E.-C.} \sur{Aschenauer}}\equalspeak{Invited speaker}
\author[35]{\fnm{M.} \sur{Balandat}}\equalspeak{Invited speaker}
\author[38]{\fnm{M.} \sur{Battaglieri}}\equalconv{Convener}
\author[6,45]{\fnm{J.} \sur{Bernauer}}\equalconv{Convener}
\author[37]{\fnm{M.} \sur{Bond\`i}}\equalspeak{Invited speaker}
\author[34,15]{\fnm{N.} \sur{Branson}}\equalspeak{Invited speaker}
\author[29]{\fnm{T.} \sur{Britton}}\equalspeak{Invited speaker}
\author[30]{\fnm{A.} \sur{Butter}}\equalspeak{Invited speaker}
\author[56]{\fnm{I.} \sur{Chahrour}}
\author[29]{\fnm{P.} \sur{Chatagnon}}
\author[39]{\fnm{E.} \sur{Cisbani}}\equalconv{Convener}
\author[48]{\fnm{E. W.} \sur{Cline}}
\author[25]{\fnm{S.} \sur{Dash}}\equalconv{Convener}
\author[33]{\fnm{C.} \sur{Dean}}\equalspeak{Invited speaker}
\author[55]{\fnm{W.} \sur{Deconinck}}\equalconv{Convener}
\author[6,4]{\fnm{A.} \sur{Deshpande}}
\author[29]{\fnm{M.} \sur{Diefenthaler}}\equalconv{Convener}\equalspeak{Invited Speaker}
\author[29]{\fnm{R.} \sur{Ent}}\equalspeak{Invited speaker}
\corrauthor[64,29]{\fnm{C.} \sur{Fanelli}}\equalhack{Hackathon Organizer}\equaltut{Editor}\equalcorr{Corresponding author}
\author[10]{\fnm{M.} \sur{Finger}}
\author[10]{\fnm{M.} \sur{Finger, Jr.}}
\author[5]{\fnm{E.} \sur{Fol}}\equalspeak{Invited speaker}
\author[29]{\fnm{S.} \sur{Furletov}}\equalspeak{Invited speaker}
\author[3]{\fnm{Y.} \sur{Gao}}
\author[64,57]{\fnm{J.} \sur{Giroux}}\equalspeak{Invited speaker}\equalhack{Hackathon}
\author[59]{\fnm{N. C.} \sur{Gunawardhana Waduge}}
\author[9,12]{\fnm{R.} \sur{Harish}}
\author[55,58]{\fnm{O.} \sur{Hassan}}
\author[9]{\fnm{P. L.} \sur{Hegde}}
\author[17]{\fnm{R. J.} \sur{Hern\'andez-Pinto}}
\author[27]{\fnm{A.} \sur{Hiller Blin}}\equalspeak{Invited speaker}
\author[49]{\fnm{T.} \sur{Horn}}\equaltut{Editor}
\author[3]{\fnm{J.} \sur{Huang}}\equalspeak{Invited speaker}
\author[22,29]{\fnm{D.} \sur{Jayakodige}}
\author[41]{\fnm{B.} \sur{Joo}}\equalspeak{Invited speaker}
\author[57]{\fnm{M.} \sur{Junaid}}
\author[32]{\fnm{P.} \sur{Karande}}
\author[8]{\fnm{B.} \sur{Kriesten}}
\author[62]{\fnm{R.} \sur{Kunnawalkam Elayavalli}}\equalspeak{Invited speaker}
\author[3]{\fnm{M.} \sur{Lin}}
\author[41]{\fnm{F.} \sur{Liu}}\equalspeak{Invited speaker}
\author[59]{\fnm{S.} \sur{Liuti}}\equalspeak{Invited speaker}
\author[16]{\fnm{G.} \sur{Matousek}}
\author[16]{\fnm{M.} \sur{McEneaney}}\equalspeak{Invited speaker}
\author[29]{\fnm{D.} \sur{McSpadden}}\equalhack{Hackathon}
\author[53]{\fnm{T.} \sur{Menzo}}\equalspeak{Invited speaker}
\author[18]{\fnm{T.} \sur{Miceli}}\equalspeak{Invited speaker}
\author[31]{\fnm{V.} \sur{Mikuni}}\equalspeak{Invited speaker}
\author[54]{\fnm{R.} \sur{Montgomery}}\equalconv{Convener}
\author[31]{\fnm{B.} \sur{Nachman}}\equalconv{Convener}\equalspeak{Invited Speaker}
\author[36]{\fnm{R. R.} \sur{Nair}}
\author[64]{\fnm{J.} \sur{Niestroy}}
\author[17]{\fnm{S. A.} \sur{Ochoa Oregon}}
\author[63]{\fnm{J.} \sur{Oleniacz}} 
\author[3]{\fnm{J. D.} \sur{Osborn}}\equalconv{Convener}
\author[19]{\fnm{C.} \sur{Paudel}}
\author[16]{\fnm{C.} \sur{Pecar}}\equalspeak{Invited speaker}
\author[1]{\fnm{C.} \sur{Peng}}\equalspeak{Invited speaker}
\author[18]{\fnm{G. N.} \sur{Perdue}}\equalconv{Convener}
\author[11,29]{\fnm{W.} \sur{Phelps}}\equalspeak{Invited speaker}
\author[3]{\fnm{M. L.} \sur{Purschke}}
\author[29]{\fnm{K.} \sur{Rajput}}\equalspeak{Invited speaker}\equalhack{Hackathon}
\author[31]{\fnm{Y.} \sur{Ren}}\equalconv{Convener}\equalspeak{Invited speaker}
\author[17]{\fnm{D. F.} \sur{Renteria-Estrada}}
\author[2]{\fnm{D.} \sur{Richford}}
\author[40,23]{\fnm{B. J.} \sur{Roy}}
\author[47]{\fnm{D.} \sur{Roy}}
\author[29]{\fnm{N.} \sur{Sato}}\equalspeak{Invited speaker}
\author[29,42]{\fnm{T.} \sur{Satogata}}\equalspeak{Invited speaker}
\author[13,21]{\fnm{G.} \sur{Sborlini}}
\author[29]{\fnm{M.} \sur{Schram}}\equalconv{Convener}
\author[46]{\fnm{D.} \sur{Shih}}\equalspeak{Invited speaker}
\author[44]{\fnm{J.} \sur{Singh}}
\author[4,7]{\fnm{R.} \sur{Singh}}
\author[28]{\fnm{A.} \sur{Siodmok}} 
\author[64]{\fnm{P.} \sur{Stone}}
\author[64]{\fnm{J.} \sur{Stevens}}\equalconv{Convener}
\author[64]{\fnm{L.} \sur{Suarez}}
\author[57]{\fnm{K.} \sur{Suresh}}\equalhack{Hackathon}
\author[20]{\fnm{A.-N.} \sur{Tawfik}}
\author[31]{\fnm{F.} \sur{Torales Acosta}}\equalspeak{Invited speaker}
\author[18]{\fnm{N.} \sur{Tran}}\equalspeak{Invited speaker}
\author[49]{\fnm{R.} \sur{Trotta}}
\author[52]{\fnm{F. J.} \sur{Twagirayezu}}
\author[54]{\fnm{R.} \sur{Tyson}}
\author[43]{\fnm{S.} \sur{Volkova}}\equalspeak{Invited speaker}
\author[29,16]{\fnm{A.} \sur{Vossen}}\equalconv{Hackathon}
\author[64]{\fnm{E.} \sur{Walter}}\equalhack{Hackathon organizer}
\author[51]{\fnm{D.} \sur{Whiteson}}\equalspeak{Invited speaker}
\author[33]{\fnm{M.} \sur{Williams}}\equalspeak{Invited speaker}
\author[55]{\fnm{S.} \sur{Wu}}
\author[60]{\fnm{N.} \sur{ Zachariou}}
\author[14,26]{\fnm{P.} \sur{Zurita}}\equalconv{Convener}

\affil[1]{\orgname{Argonne National Laboratory}, \orgaddress{\city{Lemont}, \state{IL}, \postcode{60439}, \country{USA}}}
\affil[2]{\orgname{Baruch College}, \orgaddress{\city{New York}, \state{NY}, \postcode{10010}, \country{USA}}}
\affil[3]{\orgname{Brookhaven National Lab}, \orgaddress{\city{Upton}, \state{NY}, \postcode{33973}, \country{USA}}}
\affil[4]{\orgname{Brookhaven National Laboratory}, \orgdiv{Department of Physics}, \orgaddress{\city{Upton, NY}, \postcode{11973-5000}, \country{USA}}}
\affil[5]{\orgname{CERN}, \orgdiv{Accelerators and Beam Physics Group}, \orgaddress{\city{Geneva}, \postcode{CH-1211}, \country{Switzerland}}}
\affil[6]{\orgname{Center for Frontiers in Nuclear Science, Stony Brook University}, \orgaddress{\city{Stony Brook}, \state{NY}, \postcode{11790-3800}, \country{USA}}}
\affil[7]{\orgname{Center for Frontiers in Nuclear Science, Stony Brook University}, \orgdiv{Department of Physics and Astronomy}, \orgaddress{\city{Stony Brook}, \state{NY}, \postcode{11794}, \country{USA}}}
\affil[8]{\orgname{Center for Nuclear Femtography}, \orgaddress{\city{Washington DC}, \postcode{20005}, \country{USA}}}
\affil[9]{\orgname{Central University of Karnataka}, \orgaddress{\city{Kalaburagi}, \postcode{585367}, \country{India}}}
\affil[10]{\orgname{Charles University, Faculty of Mathematics and Physics}, \orgaddress{\city{Prague}, \postcode{18000}, \country{Czech Republic}}}
\affil[11]{\orgname{Christopher Newport University}, \orgaddress{\city{Newport News}, \state{VA}, \postcode{11606}, \country{USA}}}
\affil[12]{\orgname{Deggendorf Institute of Technology}, \orgaddress{\city{Deggendorf}, \postcode{94469}, \country{Germany}}}
\affil[13]{\orgname{Departamento de F\'isica Fundamental e IUFFyM, Universidad de Salamanca}, \orgaddress{\city{Salamanca}, \postcode{37008}, \country{Spain}}}
\affil[14]{\orgname{Departamento de F\'isica Te\'orica \& IPARCOS, Universidad Complutense de Madrid}, \orgaddress{\city{ adrid}, \postcode{E-28040}, \country{Spain}}}
\affil[15]{\orgname{Drexel University}, \orgaddress{\city{Philadelphia}, \state{PA}, \postcode{19104}, \country{US}}}
\affil[16]{\orgname{Duke University}, \orgaddress{\city{Durham}, \state{NC}, \postcode{27708}, \country{USA}}}
\affil[17]{\orgname{Facultad de Ciencias F\'isico-Matem\'aticas, Universidad Aut\'onoma de Sinaloa, Ciudad Universitaria}, \city{Culiac\'an, Sinaloa} \postcode{80000}, \country{M\'exico}}
\affil[18]{\orgname{Fermilab}, \orgaddress{\city{Batavia}, \state{IL}, \postcode{60510}, \country{USA}}}
\affil[19]{\orgname{Florida International University}, \orgaddress{\city{Miami}, \state{FL}, \postcode{33199},  \country{USA}}}
\affil[20]{\orgname{Future University in Egypt}, \orgaddress{\city{New Cairo}, \postcode{11835}, \country{Egypt}}}
\affil[21]{\orgname{GFIMA - Escuela de Ciencias, Ingenier\'ia y Dise\~no, Universidad Europea de Valencia}, \orgaddress{\city{Valencia}, \postcode{46010}, \country{Spain}}}
\affil[22]{\orgname{Hampton University}, \orgaddress{\city{Hampton}, \state{VA}, \postcode{23668}, \country{USA}}}
\affil[23]{\orgname{Homi Bhabha National Institute}, \orgaddress{\city{Mumbai}, \postcode{400094}, \country{India}}}
\affil[24]{\orgname{INFN}, \orgdiv{Sezione di Roma Tor Vergata}, \orgaddress{\city{Rome}, \postcode{I-00133}, \country{Italy}}}
\affil[25]{\orgname{Indian Institute of Technology Bombay}, \orgaddress{\city{Mumbai}, \state{Maharashtra}, \postcode{400076}, \country{India}}}
\affil[26]{\orgname{Institut f\"ur Theoretische Physik, Universit\"at Regensburg}, \orgaddress{\city{Regensburg}, \postcode{D-93040}, \country{Germany}}}
\affil[27]{\orgname{Institute for Theoretical Physics, T\"{u}bingenUniversity}, \orgaddress{\city{T\"{u}bingen}, \postcode{72076}, \country{Germany}}}
\affil[28]{\orgname{Jagiellonian University}, \orgaddress{\city{Kraków}, \postcode{31-007}, \country{Poland}}}
\affil[29]{\orgname{Jefferson Lab}, \orgaddress{\city{Newport News}, \state{VA}, \postcode{293606}, \country{USA}}}
\affil[30]{\orgname{Laboratory Nuclear and High-Energy Physics}, \orgdiv{CNRS}, \orgaddress{\city{Paris}, \postcode{75005}, \country{France}}}
\affil[31]{\orgname{Lawrence Berkeley National Laboratory}, \orgaddress{\city{Berkeley}, \state{CA}, \postcode{94720}, \country{USA}}}
\affil[32]{\orgname{Lawrence Livermore National Laboratory}, \orgaddress{\city{Livermore}, \state{CA}, \postcode{94550}, \country{USA}}}
\affil[33]{\orgname{Massachusetts Institute of Technology}, \orgaddress{\city{Cambridge}, \postcode{021339}, \state{MA}, \country{USA}}}
\affil[34]{\orgname{Messiah University}, \orgaddress{\city{Mechanicsburg}, \postcode{17055}, \state{PA}, \country{US}}}
\affil[35]{\orgname{Meta}, \orgaddress{\city{Menlo Park}, \state{CA}, \postcode{935025}, \country{USA}}}
\affil[36]{\orgname{National Centre For Nuclear Research}, \orgaddress{\city{Warsaw}, \postcode{02-093}, \country{Poland}}}
\affil[37]{\orgname{National Institute for Nuclear Physics}, \orgdiv{Catania}, \orgaddress{\city{Catania}, \postcode{95125}, \country{Italy}}}
\affil[38]{\orgname{National Institute for Nuclear Physics}, \orgdiv{Genoa}, \orgaddress{\city{Genoa}, \postcode{I-16146}, \country{Italy}}}
\affil[39]{\orgname{National Institute for Nuclear Physics}, \orgdiv{Roma}, \orgaddress{\city{Rome}, \postcode{I-00185}, \country{Italy}}}
\affil[40]{\orgname{Nuclear Physics Division, Bhabha Atomic Research Centre}, \orgaddress{\city{Mumbai}, \postcode{400085}, \country{India}}}
\affil[41]{\orgname{Oak Ridge National Laboratory}, \orgaddress{\city{Oak Ridge}, \state{TN}, \postcode{37830}, \country{USA}}}
\affil[42]{\orgname{Old Dominion University}, \orgaddress{\city{Norfolk}, \state{VA}, \postcode{23529}, \country{USA}}}
\affil[43]{\orgname{Pacific Northwest National Laboratory}, \orgaddress{\city{Richland}, \state{WA}, \postcode{99354}, \country{USA}}}
\affil[44]{\orgname{Panjab University}, \orgaddress{\city{Chandigarh}, \postcode{PIN-160014}, \country{India}}}
\affil[45]{\orgname{RIKEN BNL Research Center}, \orgaddress{\city{Upton}, \state{NY}, \postcode{11973}, \country{USA}}}
\affil[46]{\orgname{Rutgers University}, \orgaddress{\city{Camden}, \state{NJ}, \postcode{08102}, \country{USA}}}
\affil[47]{\orgname{Rutgers University}, \orgaddress{\city{Piscataway}, \state{NJ}, \postcode{08854}, \country{USA}}}
\affil[48]{\orgname{Stony Brook University}, \orgaddress{\city{Stony Brook}, \state{NY}, \postcode{11794}, \country{USA}}}
\affil[49]{\orgname{The Catholic University of America}, \orgaddress{\city{Washington}, \state{DC}, \postcode{20064}, \country{USA}}}
\affil[50]{\orgname{University of California}, \orgdiv{Berkeley Institute for Data Science}, \orgaddress{\city{Berkeley},  \state{CA}, \postcode{94720}, \country{USA}}}
\affil[51]{\orgname{University of California Irvine}, \orgaddress{\city{Irvine}, \state{CA}, \postcode{92697}, \country{USA}}}
\affil[52]{\orgname{University of California Los Angeles}, \orgaddress{\city{Los Angeles}, \state{CA}, \postcode{90095}, \country{USA}}}
\affil[53]{\orgname{University of Cincinnati}, \orgaddress{\city{Cincinnati}, \state{OH}, \postcode{45221}, \country{USA}}}
\affil[54]{\orgname{University of Glasgow}, \orgaddress{\city{Glasgow}, \postcode{G12 8QQ}, \country{United Kingdom}}}
\affil[55]{\orgname{University of Manitoba}, \orgdiv{Physics}, \orgaddress{\city{Winnipeg}, \state{MB}, \postcode{R3T 2N2}, \country{Canada}}}
\affil[56]{\orgname{University of Michigan}, \orgaddress{\city{Ann Arbor}, \state{MI}, \postcode{48104}, \country{USA}}}
\affil[57]{\orgname{University of Regina}, \orgaddress{\city{Regina}, \state{SK}, \postcode{S4S0A2}, \country{Canada}}}
\affil[58]{\orgname{University of Victoria}, \orgaddress{\city{Victoria}, \postcode{V8P 5C2}, \country{Canada}}}
\affil[59]{\orgname{University of Virginia}, \orgaddress{\city{Charlottesville}, \state{VA}, \postcode{22904}, \country{USA}}}
\affil[60]{\orgname{University of York}, \orgaddress{\city{York}, \postcode{YO422SF}, \country{United Kingdom}}}
\affil[61]{\orgname{Université Paris-Saclay}, \orgdiv{CNRS/IN2P3, IJCLAB}, \orgaddress{\city{Orsay}, \postcode{91400}, \country{France}}}
\affil[62]{\orgname{Vanderbilt University}, \orgaddress{\city{Nashville}, \state{TN} , \postcode{37235}, \country{USA}}}
\affil[63]{\orgname{Warsaw University of Technology}, \orgaddress{\city{Warsaw}, \postcode{00-661}, \country{Poland}}}
\affil[64]{\orgname{William \& Mary}, \orgaddress{\city{Williamsburg}, \state{VA}, \postcode{231864}, \country{USA}}}




\abstract{

The Electron-Ion Collider (EIC), a state-of-the-art facility for studying the strong force, is expected to begin commissioning its first experiments in 2028.
This is an opportune time for artificial intelligence (AI) to be included from the start at this facility and in all phases that lead up to the experiments. 
%
%
The second annual workshop organized by the AI4EIC working group, which recently took place, centered on exploring all current and prospective application areas of AI for the EIC. 
This workshop is not only beneficial for the EIC, but also provides valuable insights for the newly established \epic collaboration at EIC.
This paper summarizes the different activities and R\&D projects covered across the sessions of the workshop and provides an overview of the goals, approaches and strategies regarding AI/ML in the EIC community, as well as cutting-edge techniques currently studied in other experiments.

}






\maketitle

\section{Introduction}\label{sec:intro}

\begin{figure}[!] 
    \centering
    \includegraphics[trim={1cm 1.5cm 2cm 2.5cm},clip,width=0.49\textwidth]{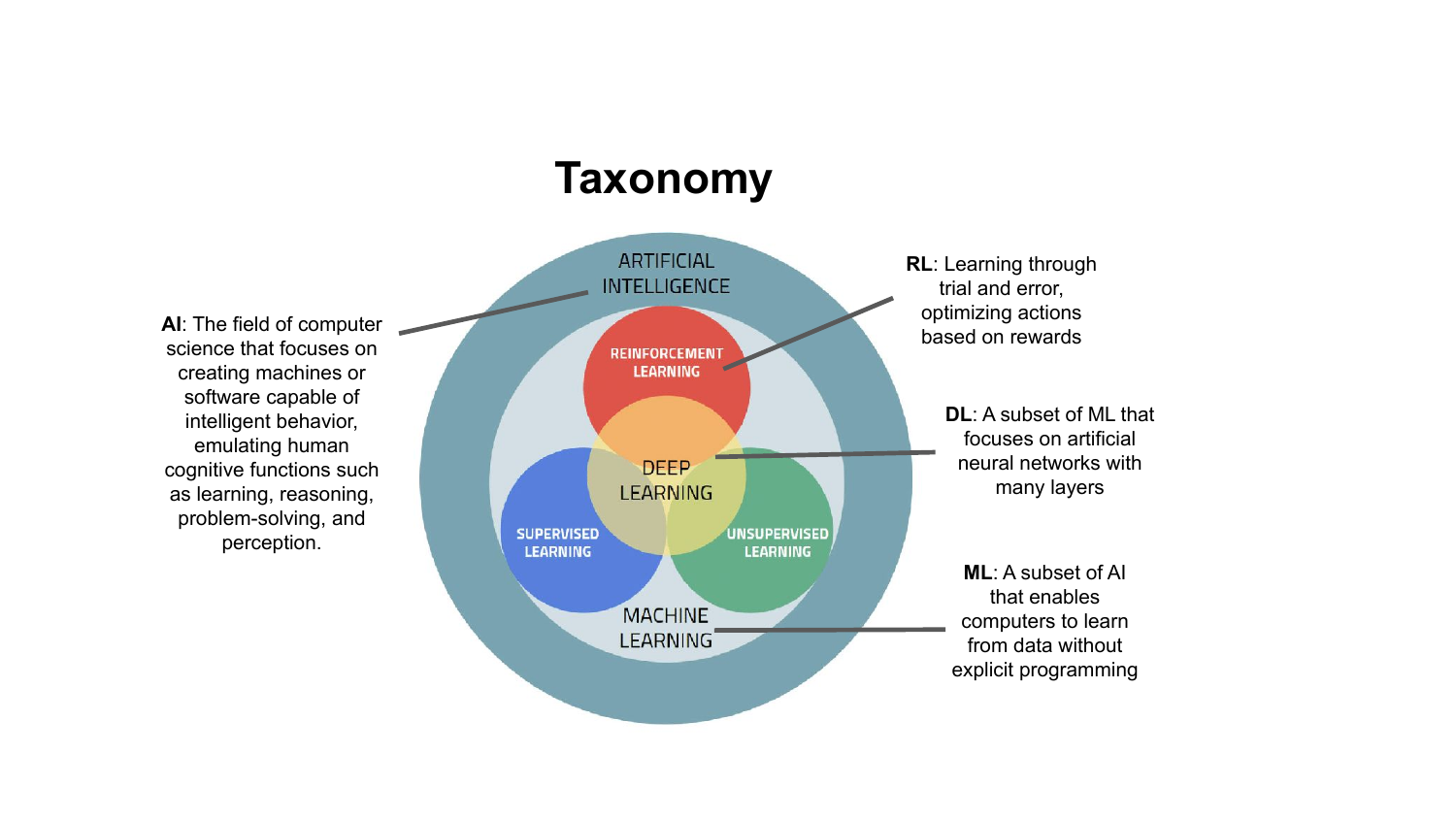}
    \caption{\textbf{Taxonomy:} A diagrammatic representation of artificial intelligence, machine learning, and deep learning is provided to familiarize readers with the corresponding acronyms utilized in the text.
    \label{fig:taxonomy}
    }
\end{figure}

In October 2022, the second workshop on Artificial Intelligence for the Electron-Ion-Collider (AI4EIC) has been held at William \& Mary. 
The workshop delved into a range of active and potential application areas of AI/ML\footnote{In this document, we follow a hierarchical taxonomy for artificial intelligence (AI), subdivided into Machine Learning (ML) and Deep Learning (DL). ML, a subset of AI, pertains to a machine's ability to deduce input-output relationships without explicit mathematical instructions. DL, a further refinement within ML, employs intricate neural networks to mimic human brain interactions, facilitating learning from unstructured inputs. See Fig. \ref{fig:taxonomy}.} for the EIC, and it was also an opportunity to showcase some of the ongoing research activities in these areas for the recently formed \epic Collaboration. 

The event also had a strong outreach and educational component with different tutorials given by experts in AI and ML from national labs, universities, and industry as well as a hackathon satellite event during the last day of the workshop.

In Table\,\ref{tab:acronyms} at the end of this document, we list many of the methods encountered in this work, with their respective acronyms.

\AtEndDocument{
\begin{table*}[!htb]
 \caption{Table of acronyms discussed in this document with description. Listed in alphabetical order.}
    \label{tab:acronyms}
    \centering
    \resizebox{0.85\textwidth}{!}{%
    \begin{tabular}{C{0.30\linewidth} P{0.70\linewidth}}
    \toprule
    \textbf{Acronym} &  
    \textbf{Brief Description} \\
    \midrule
    ACTS & A Common Tracking Software \\
    ADWIN & Adaptive Windowing \\
    AI & Artificial Intelligence \\
    AI4EIC & Artificial Intelligence for the Electron Ion Collider\\
    ASICs &  Application-Specific Integrated Circuit \\
    AWS & Amazon Web Services \\
    BNL & Berkeley National Laboratory \\
    CDC & Central Drift Chamber \\
    cMAF & Conditional Masked Autoregressive Flow \\
    cAE & Conditional Autoencoder \\
    CNN & Convolutional Neural Network \\
    CPU & Central Processing Unit \\
    DAQ & Data Acquisition \\
    DIRC & Detection of Internally Reflected Cherenkov light \\
    DIS & Deep Inelastic Scattering \\
    DL & Deep Learning \\
    DM & Diffusion Model \\
    dRICH & dual-radiation Ring Imaging Cherenkov \\
    FPGA & Field Programmable Gate Array \\
    GAN & Generative Adversarial Network \\
    GIN & Graph Isomorphism Networks \\
    GNN & Graph Neural Network \\
    GPD & Generalized Parton Distribution \\
    GPU & Graphics Processing Unit \\
    HEP & High Energy Physics \\
    JLab & Jefferson Lab\\
    LHC & Large Hadron Collider \\
    LSTM & Long-Short Term Memory \\
    MARS & Modified Multivariate value-at-risk Approximation based on Random Scalarizations \\
    MC & Monte Carlo \\
    MCEG & Monte Carlo Event Generator \\ 
    ML & Machine Learning \\
    MLP & Multi-Layer Perceptron \\    
    MLOps & Machine Learning Operations \\
    MOBO & Multi-Objective Bayesian Optimization \\
    MOEA & Multi-Objective Evolutionary Algorithm \\
    MOGA & Multi-Objective Genetic Algorithm \\
    MOO & Multi-Objective Optimization \\
    MORBO & Multi-Objective Trust-Region Bayesian Optimization \\
    NF & Normalizing Flow \\
    NN & Neural Network \\
    NP & Nuclear Physics \\
    ODD & Open Data Detector \\
    PDF & Parton Distribution Function \\
    PID & Particle Identification \\
    pQCD & Perturbative Quantum Chromodynamics \\
    QCD & Quantum Chromodynamics\\
    QCF & Quantum Correlation Function\\
    SI-DIS & Semi-Inclusive Deep Inelastic Scattering \\    
    SRF & Super-conducting Radio Frequency \\
    SRO & Streaming Readout \\
    sWAE & Sliced-Wasserstein Auto-encoder \\
    VAE & Variational Auto-encoder \\
    VLAD & Vectors of Locally Aggregated Descriptors \\
    
    \bottomrule
     \end{tabular}
     }
\end{table*}
}

\begin{figure}[!] 
    \centering
    \includegraphics[trim={0cm 0cm 0cm 0cm},clip,width=0.485\textwidth]{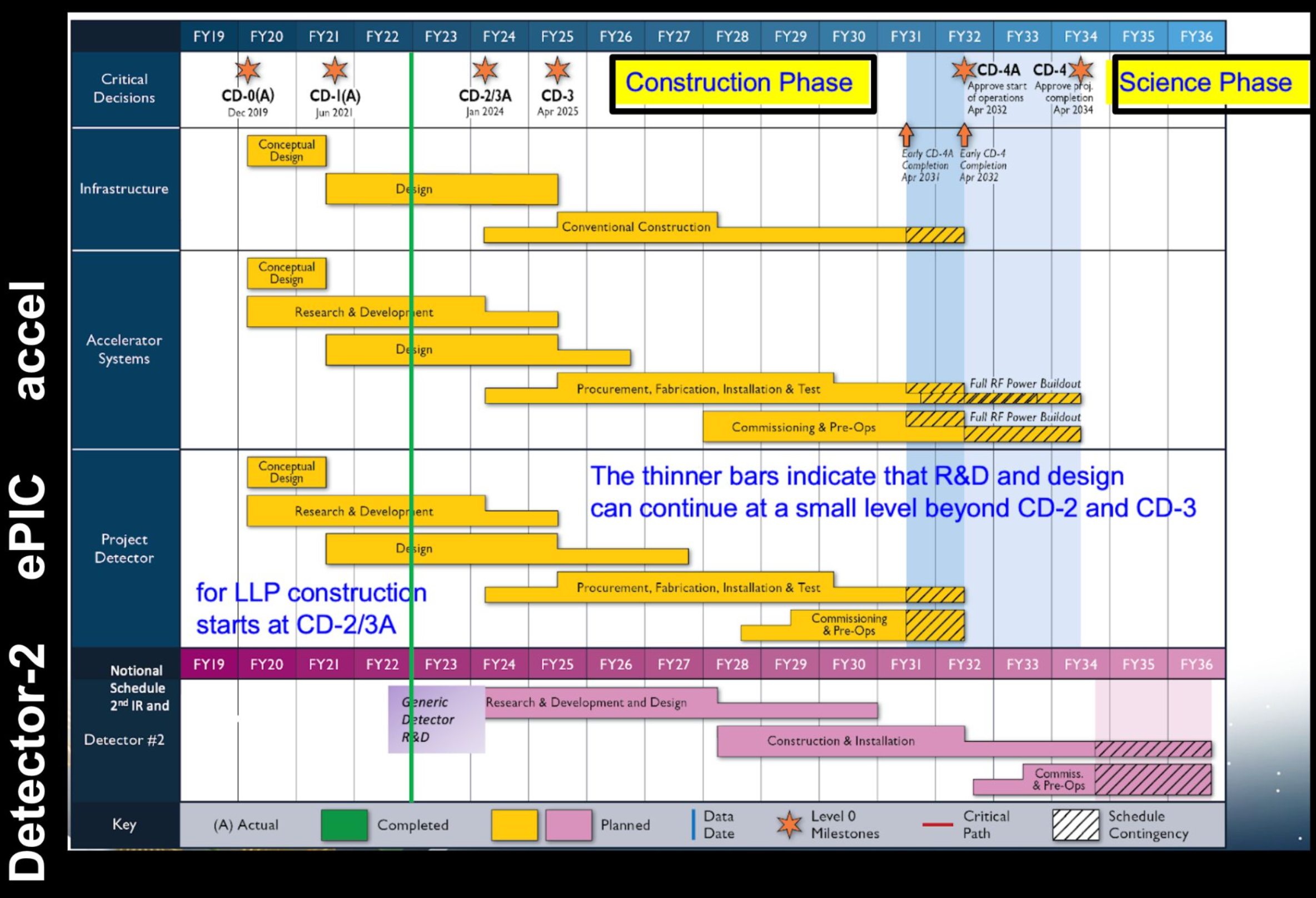}
    \caption{\textbf{EIC schedule:} the Gantt chart represents different phases (design, construction, science) for accelerator, the \epic experiment, and a potential detector-2 at EIC. Image taken from \cite{ent_ai4eic} and presented in October 2022.
    \label{fig:eic_schedule}
    }
\end{figure}

As discussed in the EIC Yellow Report \cite{khalek2021science} and as further deepened during the AI4EIC workshops, AI/ML will permeate all phases of the EIC schedule (shown in Fig. \ref{fig:eic_schedule}), and will involve accelerator and detector activities. 

The second AI4EIC workshop broadened the scope of its predecessor. While the initial workshop was centered on experimental applications for accelerators and detectors, the subsequent meeting pivoted towards the EIC detectors program, emphasizing 
 applications and fostering linkages between theoretical and experimental aspects.

The workshop was structured with the following sessions: AI/ML for Design, Experiment/Theory Connections, Reconstruction and Particle Identification (PID), AI/ML Infrastructure and Frontiers, and AI/ML in Streaming Readout (SRO). Interwoven throughout the workshop were comprehensive tutorials delivered by seasoned experts from academia, industry, and national labs.

This document is organized as follows:
\begin{itemize}
\item In Sec. \ref{sec:design}, we delve into discussions from the design session.
\item In Sec. \ref{sec:theory_exp}, we underscore the interplay between theory and experiment through AI/ML applications.
\item In Sec. \ref{sec:reco_pid}, we discuss recent advances in reconstruction and particle identification, emphasizing their applications to the EIC case.
\item In Sec. \ref{sec:infrastructure}, we detail the infrastructure solutions required for transitioning from prototype to production environments. We also address the stimulating panel discussion on AI/ML frontiers, which could shape EIC science in the coming years.
\item Sec. \ref{sec:sro} focuses on the potential of integrating AI/ML within a streaming readout data processing environment, prompting a convergence between offline and online analyses.
\item Sec. \ref{sec:community} highlights community efforts, including tutorials and a hackathon, that were conducted during the AI4EIC workshop week.
\end{itemize}

Concluding our report, Sec. \ref{sec:conclusions} encapsulates our findings and conclusions.


\section{Design of EIC}\label{sec:design}



The development of innovative experimental equipment at the EIC is skillfully leveraging cutting-edge algorithmic advancements within the dynamic landscape of AI-inspired methodologies. Throughout the instrumentation design process, decisions are made with the primary objective of optimizing performance, while thoroughly considering all project limitations and constraints.

Fundamentally, the design evolves into a meticulous optimization process of a multiparameter system, characterized either through Monte Carlo (MC) simulation or by analytical models, which are corroborated by existing experimental data and specific test results. At the EIC, accelerators and spectrometers represent complex systems, and their respective performances are optimized individually, while acknowledging their interconnected requisites. Ideally, these systems should be optimized concurrently, but current practices haven't reached this stage.

In the design session, the presenters provided a comprehensive summary of recent advancements in the application of AI-based methods to the definition and design of both spectrometer components and accelerators, encapsulating a brief overview of AI-assisted operations. The following points were emphasized:
(i) The various sub-detectors within the spectrometer should no longer be approached individually, as was the norm previously, a practice largely due to the diversity in specialized expertise and established work routines. Instead, a holistic perspective that considers all sub-detectors as a unified whole should be adopted.
(ii) Design is fundamentally a Multi-Objective Optimization (MOO) process, characterized by numerous parameters that define the system under design and several potentially conflicting objectives that need to be optimized concurrently, subject to constraints. The balancing act between optimizing objectives and adhering to constraints typically necessitates considerable computational effort and time.

\textbf{The EIC could spearhead the application of AI/ML to assist the design of large scale experiments, starting with the first detector, ePIC, and potentially extending to a second detector planned for the coming years}. Considering the ongoing AI revolution, the discussion surrounding the use of AI/ML to aid the design of these experiments is particularly relevant and timely, as their design phase is currently underway.

\begin{figure}[!ht] 
    \centering
    \includegraphics[trim={0.5cm 0cm 0cm 1.5cm},clip,width=0.485\textwidth]{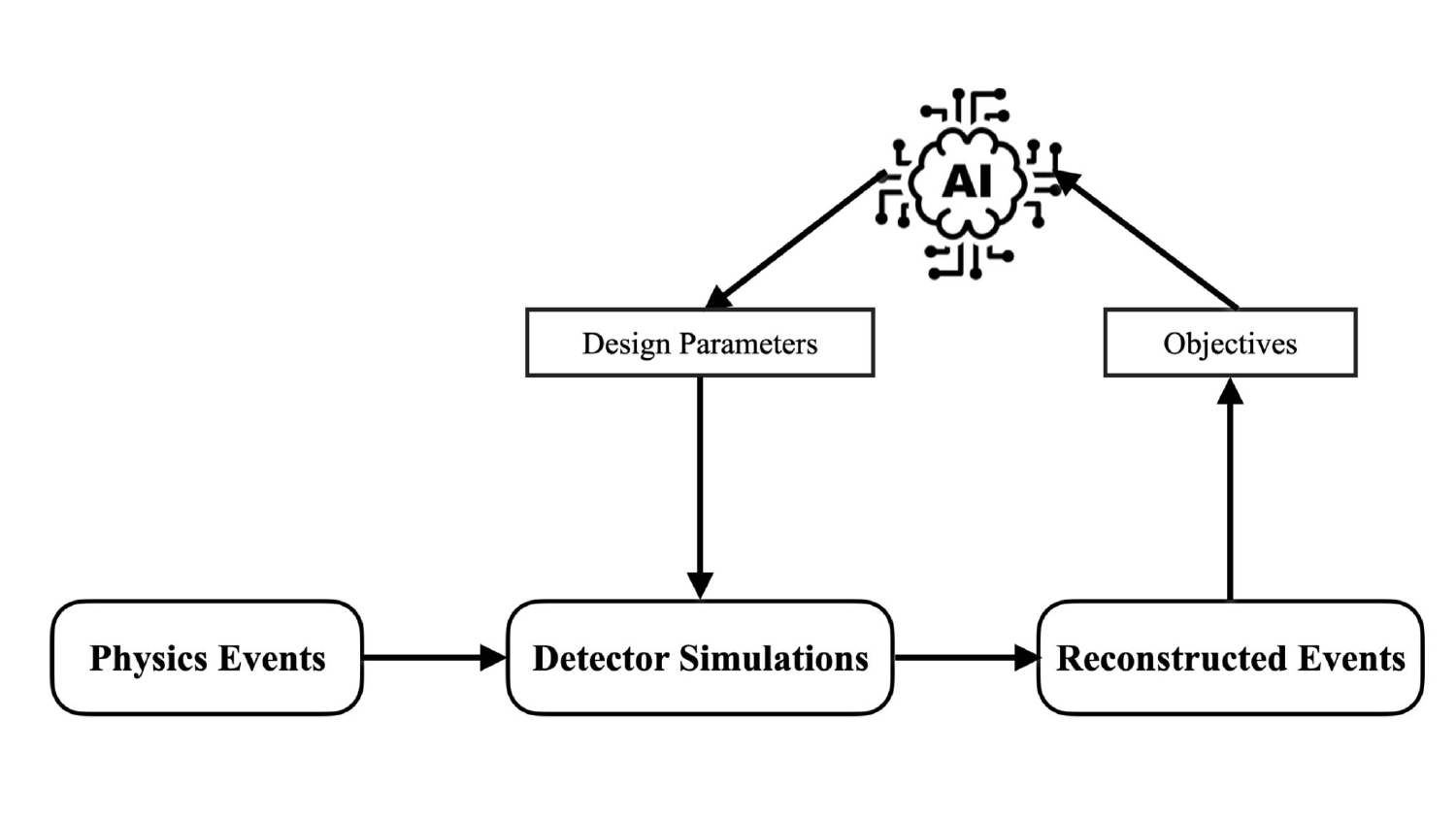}
    \caption{\textbf{AI-assisted detector design:} Flowchart of the main steps characterizing detector design optimization. Image taken from \cite{design:tracking}.
    \label{fig:ai_design}
    }
\end{figure}

A typical workflow for detector design is displayed in Fig. \ref{fig:ai_design}
%
An emerging and efficacious strategy to alleviate the computational demands of design optimization is the utilization of Parallel Bayesian Optimization. This method, which focuses on vector-based black-box functions with Expected Hypervolume Improvement \cite{design:ehi}, promises superior sample-efficiency. It accomplishes this by identifying the Pareto frontier (optimal solutions) as the most effective trade-offs.
The implementation of this approach is made simpler through the use of existing open-source libraries. These include BoTorch~\cite{design:botorch}, a Bayesian Optimization library built on PyTorch, and Ax~\cite{design:ax}, an Adaptive Experimentation Platform, which provides higher-level APIs as well as scheduling, storage, and orchestration capabilities.

The primary constraints of MOO and the measures to address them are mainly centered around four aspects:
Firstly, the issue of scalability: The model fitting process, typically utilizing a Gaussian Process for the probabilistic surrogate model, escalates at a rate of $O(n^3)$ where $n$ represents the number of data points. The quality of the model and its statistical efficiency degrade with an increase in parameters. Additionally, the hypervolume of the configuration space is super-polynomial in relation to the number of objectives. However, there are promising approaches, such as one based on a sparsity-inducing prior and Markov Chain Monte Carlo inference, designed to address high-dimensional problems where a few parameters exert a significant influence~\cite{design:saasbo}.
Secondly, the region of interest: The efficiency of the model can be improved by defining appropriate parameters like objective thresholds, in the regions of interest in the objective functions. To this end, a system is currently being developed known as MORBO (Multi-Objective 
Trust-Region Bayesian Optimization)~\cite{design:morbo}. The aim of MORBO is to increase efficiency scaling for many evaluation points by optimizing various parts of the global Pareto frontier simultaneously using a coordinated set of local trust regions.
Thirdly, the issue of noise: The model needs to be designed in a way that it can handle noisy data, including intrinsic tolerances and environmental fluctuations. Incorporating this flexibility would likely lead to more realistic and robust optimization outcomes. To optimally utilize noisy data, a MARS (Modified Multivariate value-at-risk Approximation based on Random Scalarizations) approach is currently being developed~\cite{design:mars}.
Lastly, the matter of data representation: To mitigate ill-conditioned linear systems, a minimum of double precision is recommended. The handling of discrete parameters can be accomplished through probabilistic continuous reparametrization.

The latest implementation of detector design optimization at EIC~\cite{design:fanelli} draws inspiration from the successful pilot attempt on the dual-radiator RICH (dRICH)~\cite{design:dRICH}. It harnesses the power of the Multi-Objective Evolutionary Algorithm (MOEA) and Bayesian Optimization (MOBO) libraries, integrating them with the computationally intensive Geant4-based full simulations to facilitate the ECCE tracker design~\cite{design:tracking}. This framework incorporates approximately ten design free parameters and three key objectives, subject to a variety of hard and soft constraints. It also deals with the complex requirement of preventing Geant4 volume overlaps. Key facets of the objective functions include momentum and angular resolutions, as well as the efficiency of tracking reconstruction via Kalman Filtering.
The results of the optimization have been verified by comparing them with the expected baseline performance and post-hoc reconstructed physics observables, such as $D^0 \rightarrow \pi^+ K^-$ invariant mass reconstruction. The optimization is currently in the process of transitioning from the original ECCE software framework to the more advanced \epic software framework. This transition aims to expand the AI-assisted design to accommodate a larger parameter space and include multiple sub-detectors (\textit{e.g.}, tracker, PID detectors such as the dRICH, and calorimetry) in the optimization process, along with a broader set of objectives. A significant advantage of this approach is that of utilizing accurate full simulations while limiting the number of design points necessary to approximate the Pareto front in a multi-objective space.

In EIC, an alternative Machine Learning-driven approach  has been introduced~\cite{design:calochallenge} for calorimetry design application. This approach substitutes the computationally demanding Monte Carlo simulation with an efficient surrogate model. The surrogate model utilizes generative frameworks such as Generative Adversarial Networks (GANs), Variational Autoencoders (VAEs), and Normalizing Flows (NFs) \cite{design:atlas}.\footnote{For additional details on surrogate models, please refer to Sec. \ref{sec:theory_exp}.}
This results in a differentiable simulation, in which minor perturbations are approximated using a first-order Taylor expansion. In the final step, the optimal detector parameters are identified through a Gradient Descent optimization process, assuming the use of suitable metrics. 

\begin{figure}[!ht] 
    \centering
    \includegraphics[trim={0.5cm 0cm 0cm 0cm},clip,width=0.495\textwidth]{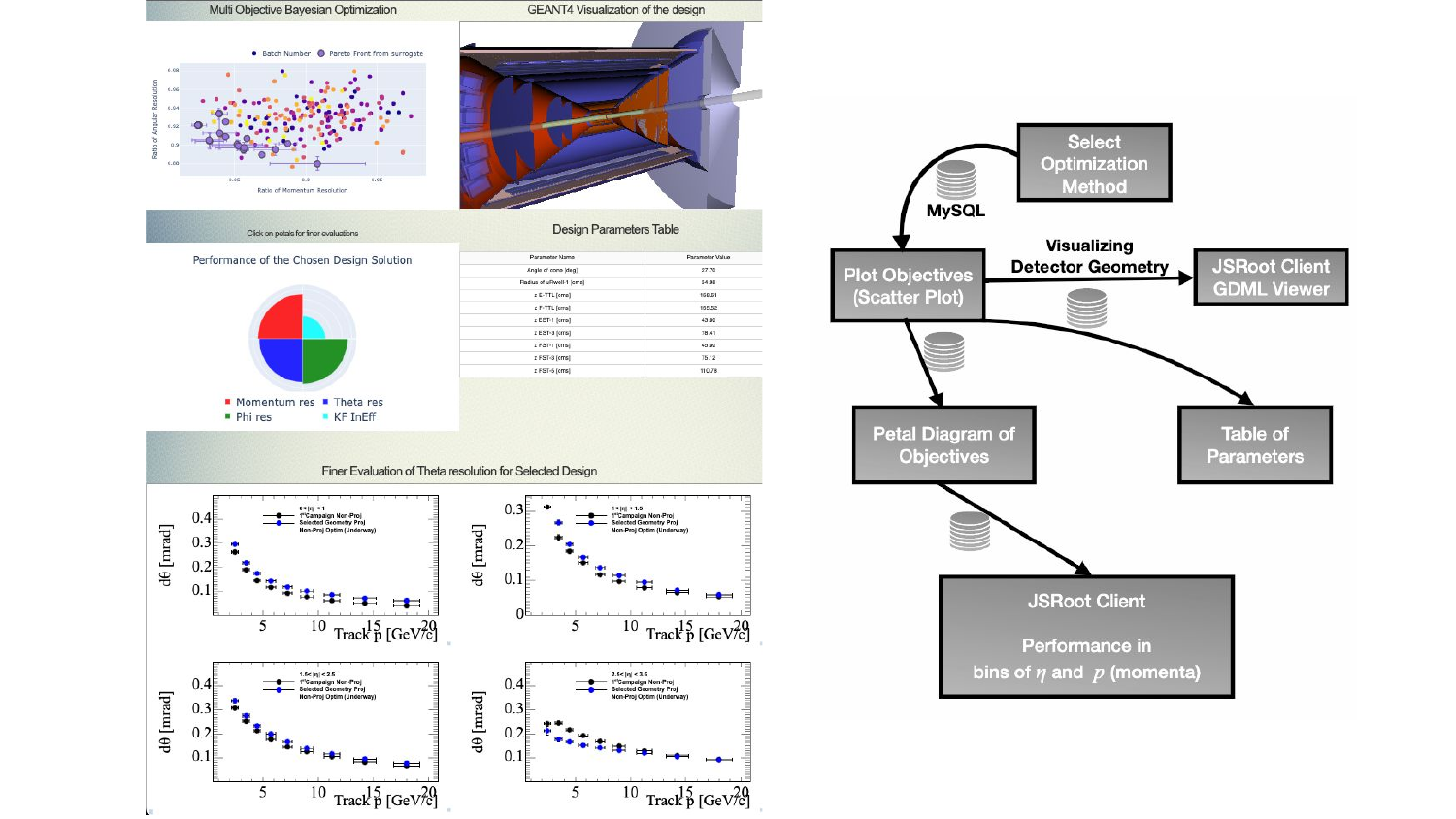}
    \caption{\textbf{Interactive Pareto front from AI-assisted design.} Left panel: interactive visualization taken from the website \cite{interactive_pareto}; right panel: A schematic of Python and JavaScript libraries facilitating result visualization utilizing advanced data science tools.
    \label{fig:interactive_pareto}
    }
\end{figure}

A significant portion of the discourse concentrated on the incorporation of cutting-edge data science tools that enable an interactive visualization of solutions within a multifaceted Pareto front. This front, which exists in a multidimensional objective space, consists of a spectrum of optimal solutions with various trade-offs between competing objectives. This ability to visualize solutions allows for a more intuitive understanding of these trade-offs and assists decision-makers in selecting the most suitable solution based on their specific preferences or constraints. An illustration of these applications, as they allow for an interactive exploration of this space, can be found in Fig. \ref{fig:interactive_pareto}. These tools, therefore, represent a critical step forward in managing the complexity of optimization problems in detector and accelerator design.
%
Beyond the progress made in automated optimization for detector design as previously mentioned, the field of accelerator science also presents fertile ground for the promising applications of optimization methodologies and AI/ML techniques.
Indeed, such methods have been integral to accelerator modelling for a number of years. Nevertheless, the application of AI/ML in accelerator science presents unique challenges. These include managing the computational complexity of the models, effectively addressing the widespread Coulomb interactions, grappling with the non-linearity inherent in many beam dynamics problems, and delicately balancing between cost, performance, technical challenges, and research-and-development efforts.
The recent Snowmass21 White Paper~\cite{design:snowmass21}, contributed to by experts from the AI/ML accelerator science community, concentrates on the priorities and strategies for accelerator modelling. It offers recommendations for the design challenges of next-generation accelerators. These include the development of an inclusive portfolio of particle accelerator and beam physics modelling tools, creation of virtual twins of accelerators, the application of advanced algorithms rooted in AI/ML and quantum computing technologies, and the establishment of efficient, scalable software frameworks.

Several promising AI/ML approaches are currently making waves in the field of accelerator physics. Firstly, the Multi-Objective Genetic Algorithm (MOGA) is a prevalent method used for the optimization of components such as Superconducting Radio Frequency (SRF) guns. Despite its regular use, MOGA still necessitates human intervention in scenarios involving parametric singularities, and the lack of harmony between the myriad approaches in use could potentially limit its overall efficiency and the fluidity of the optimization process.
Secondly, the concept of virtual or digital twins has been gaining significant traction due to its ability to generate datasets with minimal effort for the testing and training of AI/ML models, operator training, and as a natural expansion of control room online modeling. The Snowmass21 White Paper accentuates the potential of digital twins to explore broader parameter spaces. This extended exploration capacity could pave the way for the design of innovative solutions for particle acceleration in the near future.
In addition to these established methods, recent advances hold potential for future accelerator design. These include algorithmic improvements in linear algebra~\cite{design:fawzi} and non-linear/chaotic system forecasting~\cite{design:barbosa, li2022learning}, which could significantly influence accelerator surrogate models for non-linear design. However, the impact of these emergent technologies is perhaps not yet robust enough for application to the ongoing EIC accelerator design within the project's timeline.

Particle accelerator optimization poses numerous challenges, primarily stemming from the necessity to navigate non-linear, multi-objective functions that depend on thousands of dynamic machine components and settings. These factors collectively impact the design, operation, and control of particle beams, and often exceed the capacity of traditional optimization methods.
However, recent advancements have yielded promising results. For instance, decision tree-based methods have been successfully implemented to enhance Large Hadron Collider (LHC) operations, resulting in improved luminosity through more efficient beam optics control~\cite{design:ailhc}. Techniques such as Isolation and Random Forests have proven effective for instrument fault detection, as well as identification and correction of magnet errors. These applications not only uncover previously undetected hardware and electronics issues, but they also conserve operational time through early detection.
Further, autoencoder Neural Networks (NNs) have been employed to de-noise beam measurements on simulated data, leading to an anticipated improvement in measurement quality. Additionally, the use of supervised learning with linear regression models for virtual diagnostics enables the reconstruction of optics observables without direct measurements, potentially accelerating machine commissioning and mitigating the need for time-consuming measurements.
These successful applications have spurred ongoing research for the design and optics corrections in the LHC upgrade, which could potentially be adapted for EIC or inspire new advanced methodologies for collider operations.


In conclusion from the design session of the workshop, it was agreed that EIC is poised to greatly benefit from the application of AI in the control, commissioning, monitoring, and operation of accelerators and spectrometers. It was stressed that recognizing and integrating these opportunities early in the design phase is crucial.
However, it's important to underscore that the implementation of AI/ML applications in physics, or any other field, typically necessitates the incorporation of new multidisciplinary expertise, implying the potential need for the creation of specific positions within the research teams.
That said, AI/ML should not be viewed as a panacea; their application should be judiciously considered where appropriate, and especially where conventional methodologies are inadequate or underperforming.
Over the past few years, a significant challenge has been to surmount the cultural barriers that have hindered broader acceptance of AI/ML as established engineering tools in our applications. Today, there is a growing consensus that tasks involving optimization in multidimensional and multi-objective spaces can be more effectively addressed with AI/ML, rather than relying on traditional approaches. This new paradigm harnesses the power of AI/ML while simultaneously leveraging the expertise of humans in the optimization loop, which can lead to more robust and efficient solutions.

\section{Intersection between Theory and Experiment}\label{sec:theory_exp}




ML techniques have long been successfully employed as data analysis tools within the realm of experimental nuclear and particle physics. However, when it comes to theoretical or phenomenological perspectives, we've yet to fully explore the potential these techniques offer.

In the context of QCD, ML seems particularly adept at handling non-perturbative phenomena. This includes initial state parton densities (in a broader sense), as well as the final state hadronization process. These complex aspects of QCD could greatly benefit from a comprehensive application of ML.
Using QCD factorization theorems and evolution, parton distribution functions (PDFs) and other quantum correlation functions are deduced by conducting global fits on available data. This conventional method involves probing various regions in the parameter space to find the best location, given a specific objective. The ideal parameters are pinpointed by minimizing (or maximizing) a particular cost function, typically through the gradient descent algorithm.
To mitigate the influence of parametrization bias, the NNPDF collaboration introduced the application of NNs in extracting collinear proton PDFs (as referenced in \cite{Ball_2017}). Later, they extended this approach to fragmentation functions (FFs), as detailed in \cite{Bertone_2017}. 
This innovative use of NNs serves to refine our understanding and analysis of pQCD.
%
%

Inspired by their success, several groups have attempted to exploit the flexibility of the NNs to determine more complex, higher dimensional distributions \cite{Cuic_2020,Almaeen_2022}.
As a concrete example, we discuss the benefits and challenges of using NNs to extract generalized parton distributions (GPDs) as perused by the FemtoNet Collaboration \cite{Almaeen_2022}.

\begin{figure}[!]
    \centering
    \includegraphics[width=0.48\textwidth]{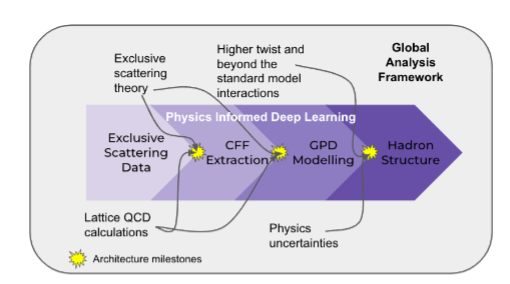}
    \caption{\textbf{FemtoNet global analysis workflow: } A Physics-Informed Deep Learning Framework that Translates Exclusive Scattering Data into Insightful Information. 
    }
    \label{fig:femto_workflow}
\end{figure}

The current knowledge of GPDs lags far behind that of collinear PDFs due to their dependence on additional kinematic variables, sparse kinematic coverage, and the overall amount of data being limited. Moreover, in the deeply virtual exclusive scattering processes of interest for GPDs, the cross-sections are written in terms of convolutions of the GPDs over one of the kinematic variables. Unlike DIS, direct access to the relevant distributions is not possible, and thus fully grasping the information enclosed in the data is a formidable task for traditional fitting methods.
The workflow of Fig. \ref{fig:femto_workflow} depicting the FemtoNet global analysis framework is a response to this challenge \cite{Almaeen_2022}. 
The main goal is to establish an unprecedented precision analysis framework to characterize the quark-gluon structure of matter. At each step of the analysis pipeline, specific physics informed deep learning architectures are implemented to ensure that the essential physics constraints are satisfied in the ML algorithms' predictions. Along the way in this analysis, there is an intentional and strategic injection of physics information from various sources such as theory, lattice QCD, and possible higher twist/beyond the standard model interactions. The pipeline culminates in the extraction of essential information on hadronic structure. 

\begin{figure}[!]
    \centering
    \includegraphics[width=0.43\textwidth]{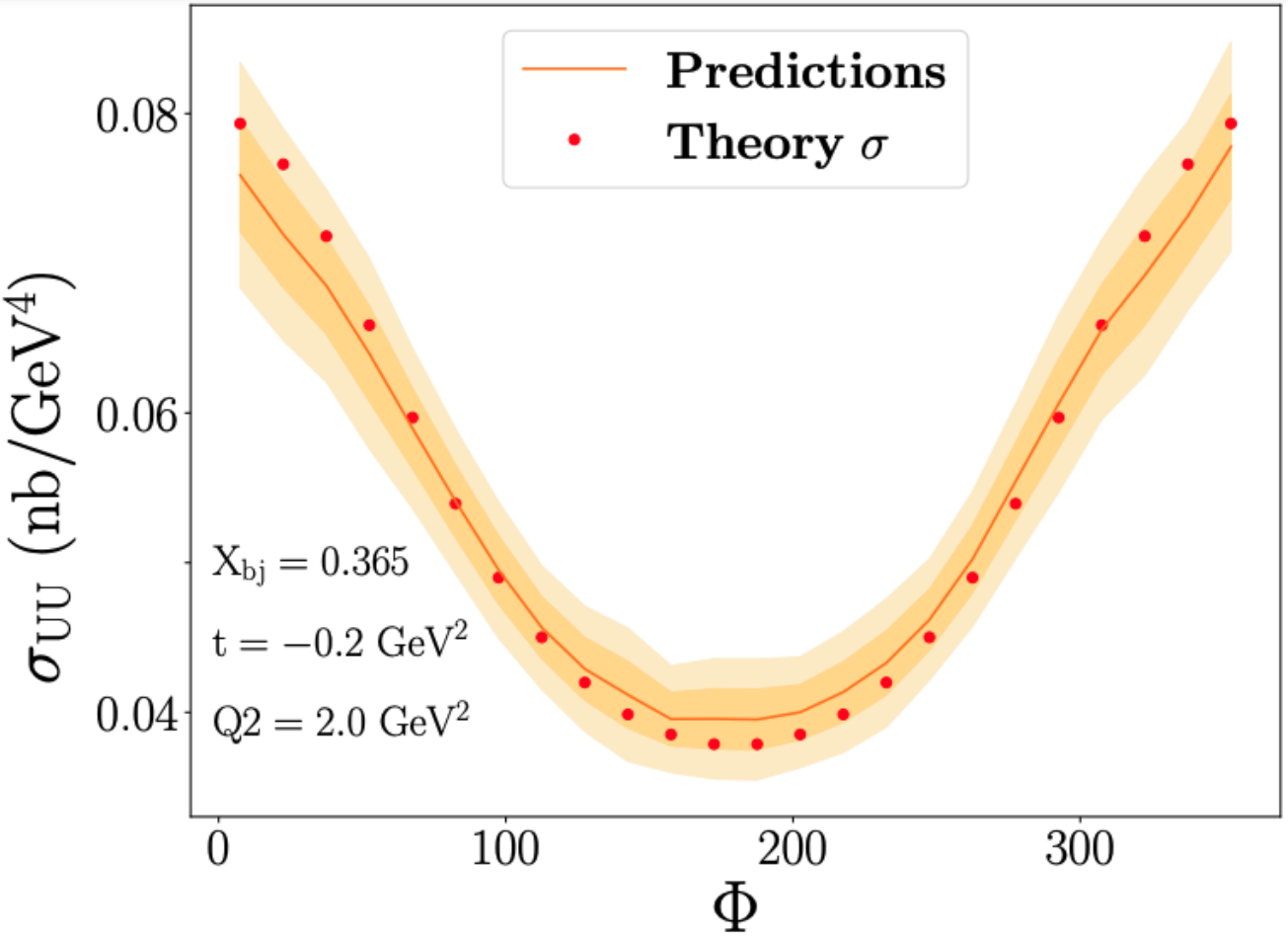}
    \caption{\textbf{FemtoNet results on DVCS cross-section modeling:} DVCS extrapolation on kinematics outside the range covered in experiment at the kinematic point $x_{Bj} = 0.365$, $t = -0.2 GeV^2$, $Q^2 = 2 GeV^2$, and $E_b = 5.75 GeV$. 
    ML Model with Angle Symmetric Constraints. 
    Figure and caption taken from \cite{Almaeen_2022}.
    }
    \label{fig:Liuti}
\end{figure}

As a first step in this analysis to determine the GPDs, the FemtoNet collaboration applies supervised learning utilizing a Multi-Layer Perceptron (MLP) complemented with regularization techniques, namely ``dropout", to prevent over-fitting \cite{Grigsby:2020auv, Almaeen_2022}. Due to the limitations in the data, a key role for generalization is played in their study by the incorporation of physical information into the neural architecture. In practice, this is done by adding terms to the loss functions, penalizing behaviors that deviate from the theoretical knowledge. 
As evidenced in Fig. \ref{fig:Liuti} and corroborated by \cite{Almaeen_2022}, the physics-informed model notably outperforms non-physics based counterparts in cross-section prediction.


%

A pivotal subject of discussion was Monte Carlo event generators (MCEGs) \cite{campbell2022event}, indispensable tools for numerical simulations and subsequent data analysis. MCEGs play a crucial role in high-energy and nuclear physics, being essential for model validation, facilitating discoveries, experimental planning, and further advancement of theories such as QCD. Their enhancement is geared towards the improvement of prediction accuracy and computational efficiency.
Traditionally, non-perturbative aspects rely on phenomenological models. These models, in turn, depend on numerous parameters that are derived from experimental data. 
An integral facet of the MCEGs is the modeling of hadronization, a transformative process wherein high-energy, color-charged quarks and gluons morph into color-neutral hadrons. 
Gaining insight into the `hadronization' process, or the mechanism by which these particles reconfigure into their final state, is crucial for establishing significant comparisons between data and theoretical predictions and improving our understanding of the hadronization process.  The high-dimensionality of these models escalates the complexity of the task at hand. Therefore, it's unsurprising that current models can't completely encapsulate data across the entire energy spectrum explored. This raises the question: could Machine Learning present a more efficacious approach?

Currently, three potential ML strategies are being considered: VAEs, GANs, and NFs \cite{Kingma_2022,Goodfellow_2014,Rezende_2016}.

VAEs have been deployed to emulate a simplified version of the Lund string model in Pythia, with the assumption of flavor and kinematics of hadron emission being independent \cite{ilten2023modeling}. In particular, the conditional sliced-Wasserstein Auto-encoder (sWAE) was trained using kinematic distributions for variables - $p'_z$ (a rescaling of $p_z$), $p_T$ extracted from Pythia's first emission events (refer to Eq. (2) of \cite{ilten2023modeling}), and specific values of string energy.
The network was tested using a unique set of string energies not included in the training set. This strategy facilitated a more accurate assessment of the network's ability to generalize across the entire phase space. To simplify the process, only pions in the final state were considered, and the performance was compared with the average Pythia output. This methodology conveniently allows the inclusion of an energy dependence in the hadronization process, if required by the data.
The first hadron emissions, which form the basis of the training, are successfully reproduced (refer to Fig.10 in \cite{ilten2023modeling}).
Comparison with the full hadronization chain (see Fig.11 in \cite{ilten2023modeling})
shows a deviation of no more than 10$\%$; such differences originate from the different treatment of the first and subsequent emissions in Pythia which is not considered in the ML approach. 
While the architecture was applied to a simplified version of the Lund string model, the results are promising and the use of ML is foreseen to be more relevant once training is done on real data, for which the hadronization is not physically accessible. 

GANs, instead, were used to learn the cluster decay of the cluster hadronization model using Herwig data \cite{Ghosh_2022}. Differing from VAE's, which learn mappings for both encoding and decoding, GANs learn only the decoding from a base distribution utilizing a discriminative loss function, comparing generations with ground truth. This was done for single $e^++e^-$ annihilation into two $\pi^0$. Despite the simplifications introduced for faster training, it was found that the method generalizes to other hadron species and, even more importantly, that the level of discrepancy with real data is similar to the one achieved with the original cluster decay model. 

NFs have been used to further improve the generation scheme, utilizing a Conditional Masked Autoregressive Flow (CMAF) \cite{Papamakarios_2018} as the generation mechanism for the kinematic \cite{Youssef_2022}. The network is conditioned on a set of hadron masses with differing initial energies. 
Contrary to earlier methods that restricted consideration to pions alone, the introduction of functional dependence via the hadron mass condition enables the generation of a range of masses in the final state \cite{Youssef_2022}. Additionally, the conditional flow adeptly captures the correlation between the $p_T$ and $p_z$ kinematic distributions.


On the experimental side of connecting theory to experiment, we have identified four major challenges. 
The first of these challenges pertains to \emph{fast simulation}, a suite of tools designed for the swift transition from particle-level predictions to detector-level observations.
Significant progress has been made in ML-based fast simulations, particularly with the advent of `surrogate models'. These models leverage various deep generative modeling approaches, including VAEs \cite{Kingma_2022}, GANs \cite{Goodfellow_2014}, NFs \cite{Rezende_2016}, and Diffusion Models (DMs) \cite{Dickstien_2015}.
Much of the community's attention has been devoted to the simulation of calorimeters, which typically form the slowest segment of the simulation stack \cite{Chekalina_2019,Rogachev_2023,Ratnikov_2023,Mikuni_2022}. Calorimeters, featuring both longitudinal and transverse segmentation, offer a high-dimensional emulation space. Despite the complexity, the latest neural network models have managed to mimic \textsc{Geant4} simulations \cite{Agostinelli_2003} with impressive accuracy \cite{krause2021caloflow}.

The second significant challenge lies in \emph{reconstruction}. Traditional shallow learning has long been employed for tasks such as momentum reconstruction and particle identification. However, the advent of DL has ushered in innovative methods that process low-level inputs in a more comprehensive manner.
%
%
Furthermore, ML continues to influence even the most foundational tasks in data analysis. The reconstruction of the kinematic variables in DIS such as Bjorken $x$ and four-momentum transfer squared $Q^2$ is being reevaluated in light of the advancements made in ML. It has been shown for inclusive DIS measurements that the reconstruction methods benefit from the application of ML-based models \cite{Diefenthaler_2022_paper, Arratia_2022}. This is now studied further for semi-inclusive DIS where it is important to precisely determine the inelasticity $y$ of the process and the azimuthal angles of the final state. 
The success of ML in these foundational areas underlines its potential for tackling complex problems in nuclear and particle physics. It also underscores the importance of developing ML techniques that are robust and versatile enough to meet the demands of a rapidly evolving scientific landscape.

The third critical challenge is tied to \emph{parameter inference}. This aspect distinctly differs from reconstruction, which focuses on deducing properties of an individual event or a single object within that event. In contrast, parameter inference is concerned with extracting physical parameters from entire datasets, thereby providing a holistic understanding of the reaction or system under consideration.
%
%
Such a nuanced analysis is invaluable in scenarios with complex or high-dimensional datasets, where conventional statistical methods may struggle. For example, recently, DL have been used in the search for exotic hadrons where production and decay parameters may be determined from models built on the underlying quantum mechanical amplitudes~\cite{Ng:2021ibr,Liu:2022uex}.  By leveraging ML, we can uncover intricate correlations and patterns that might otherwise remain hidden. This makes ML an indispensable tool for parameter inference in the modern data-centric world.

Finally, the fourth challenge pertains to \emph{cross-section inference}. For a vast array of measurements, experimental teams provide corrected differential cross-section results in a readily usable format for subsequent inferential tasks outside the scope of the originating collaboration. ML is precipitating a paradigm shift in how we execute these corrections, commonly known as deconvolution or unfolding. Cutting-edge methods have facilitated the unfolding of high-dimensional and unbinned data \cite{andreassen2020omnifold,chan2023unbinned,Andreev_2022}. This development is paramount to effectively harness EIC data, given that intricate correlations across numerous dimensions are necessary to comprehensively analyze the three-dimensional structure of the proton.

In conclusion, we highlight two frameworks that amalgamate theoretical and experimental aspects, and include uncertainty quantification:
the A(i)DAPT group \cite{Hiller_2022,alanazi2021survey}, has introduced an innovative employment of ML-based MCEG for data analysis and preservation. Its objectives encompass data compression, providing powerful interpolation tools, and the ability to unfold detector effects, enabling the acquisition of accurate vertex-level data. Additionally, the framework incorporates a GAN-based surrogate model for rapid detector folding, as demonstrated in \cite{alanazi2022machine}. Successful testing and validation of this framework, along with its potential to mitigate theory bias during the inference of event distributions, represent a significant advancement towards the reconstruction of physical observables.
The QuantOm collaboration \cite{Sato_2022}, has presented another pioneering approach that adopts a holistic strategy for global analysis, seamlessly integrating theoretical and experimental components. By employing an event-based analysis methodology, this approach capitalizes on generative models such as GANs to establish an event-level Quantum Correlation Function (QCF) inference framework. This framework provides a comprehensive and advanced perspective on 3D hadron tomography and nuclear imaging.

\section{Reconstruction and Particle Identification}\label{sec:reco_pid}



PID and reconstruction are crucial components of physics analyses at EIC. 

The integration of AI and ML in these domains is witnessing a rapid surge, providing promising opportunities for performance enhancement and comprehensive utilization of detector information beyond conventional techniques in these fields.

This session featured eight insightful talks, meticulously selected for their direct relevance to ongoing EIC endeavors and their complementarity to additional inputs we collected during the first AI4EIC workshop, whose contributions are also encapsulated in this section. 
The content from the second workshop broadly covered four pivotal subjects:
(i) Reconstruction/PID; (ii) Tracking; (iii) Jet Classification; and (iv) Domain Adaptation and Data-Driven Methods.


At the workshop, we had four contributions to reconstruction/PID via calorimeter responses. 
The first study focused on the use of interpretable networks for lepton identification amidst jet-induced backgrounds \cite{Whiteson_2022}. This paper underlined how deep learning, like Convolutional Neural Networks (CNN), can uncover overlooked low-level image data and isolate novel high-level features, outperforming traditional high-level feature physics models.
The second contribution was about muon identification with DL \cite{Phelps_2022}, showing that modern deep-learning architectures that efficiently combine the information coming from the tracking and calorimetry sub-systems can learn how to distinguish charged $\mu$’s from charged $\pi$’s, the latter representing the main background source.  
Our third contribution coupled the reconstruction of shower profiles within a hybrid barrel calorimeter, integrating sandwiched layers of monolithic silicon sensors Astropix and Pb/ScFi fibers, with a CNN for PID assessment of these profiles \cite{Peng_2022}. This synergy between hardware and deep learning enable superior e/$\pi$ separation, precise $\gamma$ and $\pi^{0}$ differentiation, radiative $\gamma$ tagging, and low-energy $\mu$ identification, impacting multiple areas of the EIC physics, such as DIS, deeply virtual Compton scattering, QED internal corrections, J/$\psi$ and timelike Compton scattering.
\begin{figure}
    \centering
    \includegraphics[width=0.45\textwidth]{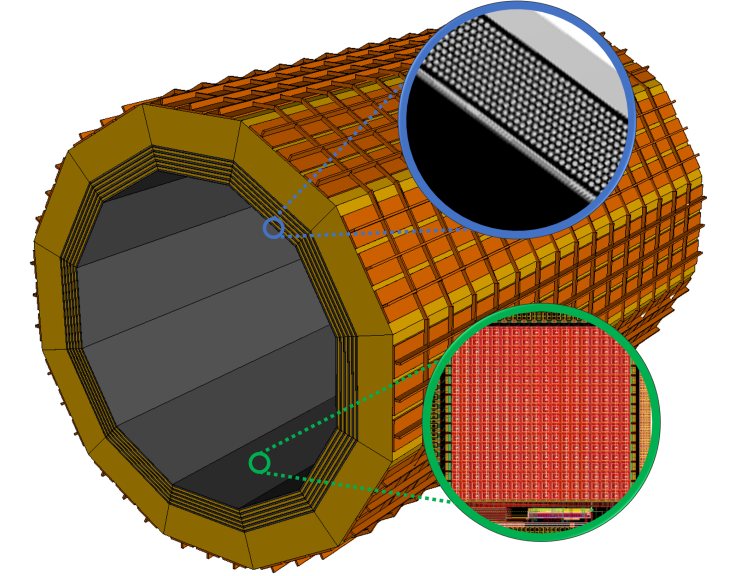}
    \includegraphics[width=0.45\textwidth]{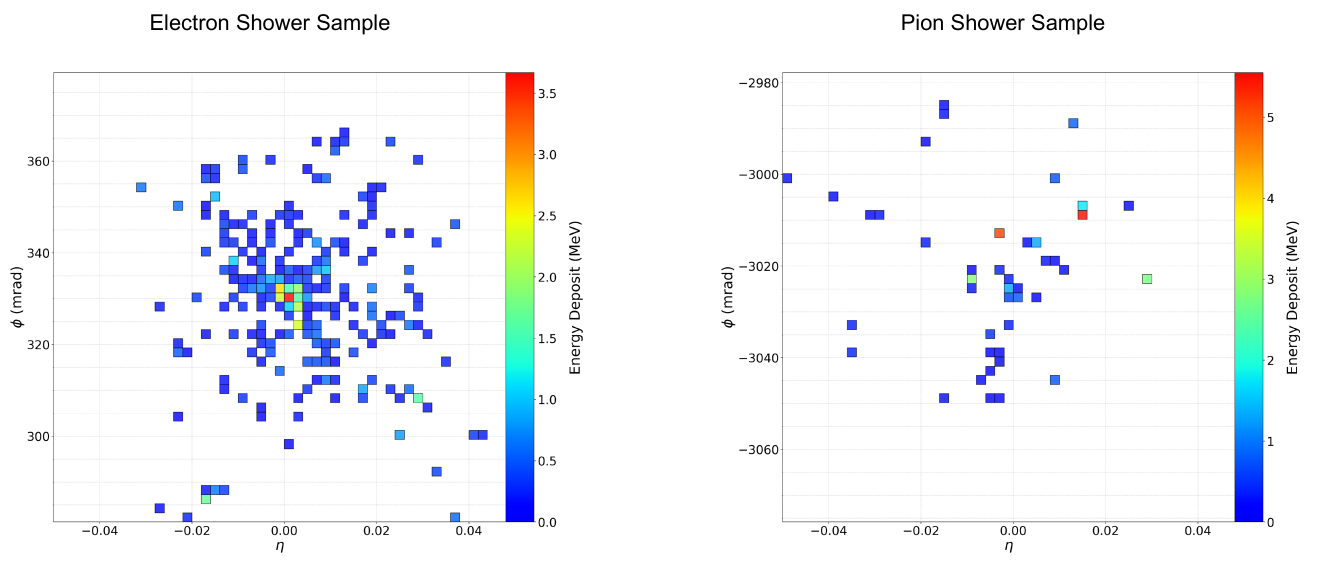}
    \caption{\textbf{(top) ECal hybrid concept:} the barrel hybrid electromagnetic calorimeter concept for EIC. More details can be found in \cite{apadula2022monolithic}; \textbf{(bottom) Projection of Showers in the ECal:} Shower projections of electrons (left) and pions (right) as a function of $\psi$ and $\eta$. Energy deposition in the pixelated array is represented via color, commonly occupying the channel axis in vision-based neural networks.}
    \label{fig:Chao_EDep}
\end{figure}
The fourth study delved into the application of ML for pixelated calorimetry, specifically for cluster separation in the electron endcap \cite{Branson_2022}. Different Artificial Intelligence/Machine Learning (AI/ML) techniques were evaluated, with a particular focus on using VAEs. VAEs were leveraged on a full-scale calorimeter to condense clusters into single points representing their total energies.


In the preceding workshop, we emphasized the vital role of Cherenkov detectors for Electron-Ion Collider (EIC) experiments \cite{Joosten_2021, Fanelli_2021}. As the primary component of the PID for the ePIC detector, these units are equipped with a Ring Imaging Cherenkov (RICH) detector in the electron endcap, a dRICH in the hadronic endcap, and a Detection of Internally Reflected Cherenkov light (DIRC) also in the hadronic endcap. This setup ensures superior PID capabilities across a broad range of the particle phase space \cite{khalek2021science}.
When it comes to Cherenkov detectors, there are two challenging areas. The first pertains to the simulation of these detectors, which typically demands significant computational resources. This is because the process involves tracking a substantial number of photons across complex surfaces, as illustrated in the contribution \cite{Joosten_2021}.
The second area of challenge lies in the reconstruction process, specifically in recognizing patterns of sparse ring images amid noisy conditions. This complexity is further exacerbated in the context of DIRC detectors due to the intricate ring topologies.
In addressing these challenges, advancements in ML and DL present considerable promise for enriching the state of the art in both reconstruction and PID in relation to Cherenkov detectors \cite{fanelli2020machine}. As discussed in \cite{Fanelli_2021,fanelli2022artificial}, algorithms such as DeepRICH \cite{fanelli2020deeprich} leverage generative models to offer rapid, accurate simulations. Additionally, as demonstrated in the context of the DIRC detector, these algorithms are capable of reconstructing intricate hit patterns, with performance on par with traditional reconstruction methods, but at a significant speed—roughly four orders of magnitude faster during inference time on a Graphics Processing Unit (GPU).
Furthermore, we highlighted the substantial opportunities presented by ML/DL applications, which enable learning at the event level as opposed to the particle level. This approach not only leverages the additional information characterizing each event, but also effectively manages the simultaneous detection of multiple particles within the detectors. This shift in focus coupled with the possibility to train these models on high-purity real data, can lead to deeper understanding of the detector response.



In the realm of jet classification (for a comprehensive review on this topic, the reader may refer to, \textit{e.g.}, \cite{boehnlein2022colloquium, feickert2021living}), a novel approach, JetVLAD \cite{Raghav_2022,bielvcikova2021identifying}, was presented. JetVLAD employs vectors of locally aggregated descriptors (VLAD) to tag heavy-flavor jets, proving instrumental for examining jet interactions with the quark-gluon plasma (QGP) created in high-energy heavy ion collisions. Such interactions are crucial for understanding partonic energy loss within the QGP medium.
The JetVLAD architecture, mirroring the ResNet model family \cite{he2016deep}, uses residual blocks with batch normalization to simplify learning. The model's width was designed to match the output of the NetVLAD layer \cite{arandjelovic2016netvlad}, a CNN architecture created for weakly supervised place recognition.
This innovative approach efficiently identifies jet flavor, enabling the analysis of mass dependence in jet-QGP interactions, and sets the stage for high-purity heavy-flavor measurements in contemporary and forthcoming collider experiments.
In our inaugural workshop, we explored the potential of AI/ML in enhancing heavy-flavor and jet tagging in EIC experiments \cite{Sekula_2021}. This insightful presentation emphasized the critical role of merging low-level and high-level track/calorimeter data for the efficient identification of jets or heavy flavor states, and showcased several effective examples of such implementations drawn from LHC studies, like the possibility of simultaneous estimation of b jet energy and resolution \cite{sirunyan2020deep}.
In the discussion concerning AI/ML applications for jets, it was mentioned a manuscript published concurrently with the workshop in October 2022 \cite{lee2023machine}. Notably, this work delves in the usage of out-of-jet radiation information, incorporates infrared jet flavor definition for handling non-perturbative QCD effects, and underscores the potential for training such deep learning models with real data. 
The integration of deep learning for jet analysis could profoundly impact EIC research by reinforcing constraints on transverse momentum-dependent PDFs, augmenting sensitivity to transverse single spin asymmetry, and elucidating cold nuclear matter effects. More comprehensive information can be gleaned from the manuscript \cite{lee2023machine}.


Regarding AI/ML for tracking at the EIC, we extended the discussion initiated in the first workshop \cite{Gagnon_2021,gagnon2022machine}, taking cues from the forthcoming upgrade of the LHC to the HL-LHC. Despite the proficiency of existing track reconstruction algorithms based on Kalman filters, they encounter scaling issues with increased data volumes. This necessitates active research into new or enhanced algorithms, involving accelerated hardware application of existing Kalman filters, the integration of ML techniques, and the creation of complete ML-pipelines for tracking like those proposed by the Exa.TrkX project \cite{tsaris2018hep}. Also, A Common Tracking Software (ACTS), a new algorithm test bed for track reconstruction research, was highlighted \cite{Ai:2021ghi}, and in the second AI4EIC workshop, we decide to delve deeper into this topic. ACTS is an agnostic, open-source tracking toolkit \cite{Allaire_2022}. Written in C++, ACTS streamlines the entire track fitting process and provides an example framework with Python bindings. Its utilization spans various experiments, like ATLAS, ALICE, sPHENIX, and EIC studies. ACTS serves as a comprehensive tool for developing and testing new ML-based tracking algorithms, making it crucial for current EIC advancements. It also offers an open data detector (ODD) for algorithm benchmarking and ML tracking tests. Noteworthy tools available in ACTS include hashing for hits selection, parameter auto-tuning, and Graph Neural Network (GNN) for track finding.
Regarding GNN for tracking, we had discussed the deployment of GNN in a streamlined pipeline for trigger-background event classification in both sPHENIX and EIC and its implementation on Field Programmable Gate Arrays (FPGAs) \cite{Dantong_2021,xuan2022high}. This subject is expanded further in Sec. \ref{sec:sro}.

\begin{figure}[h!]
    \centering
    \includegraphics[width=0.23\textwidth]{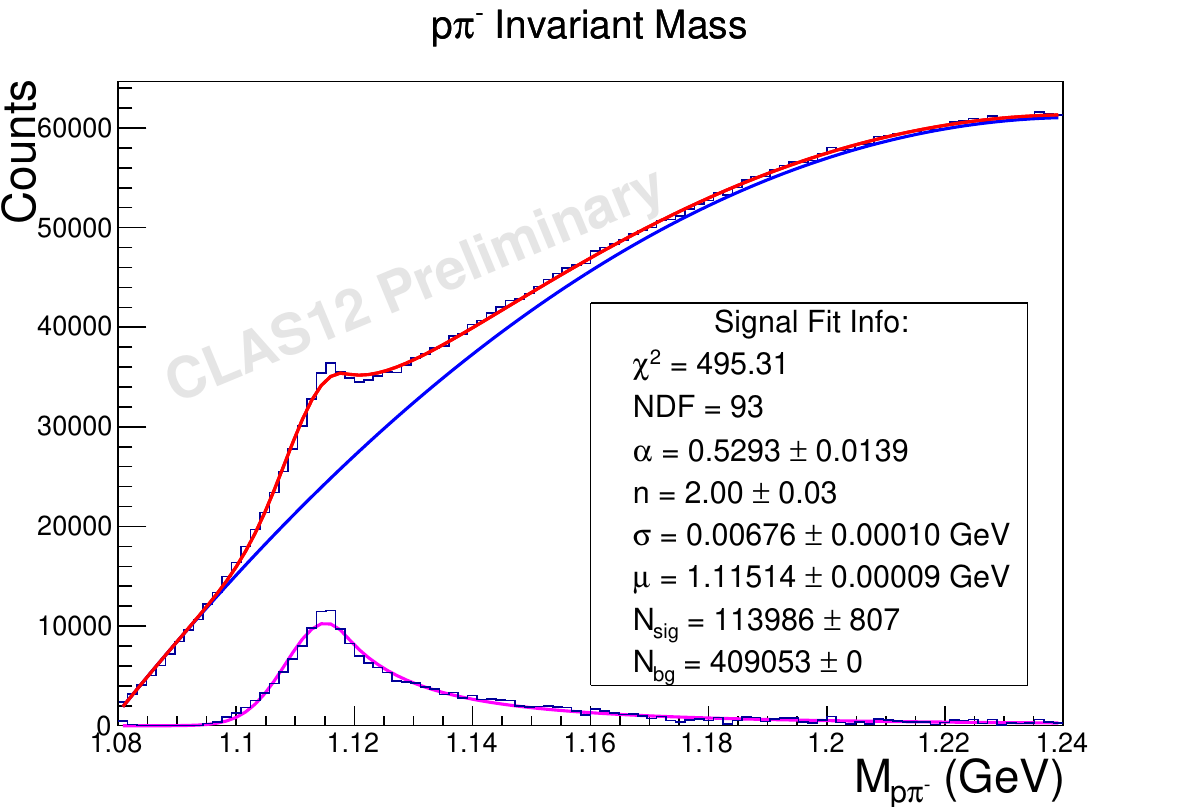}
    \includegraphics[width=0.23\textwidth]{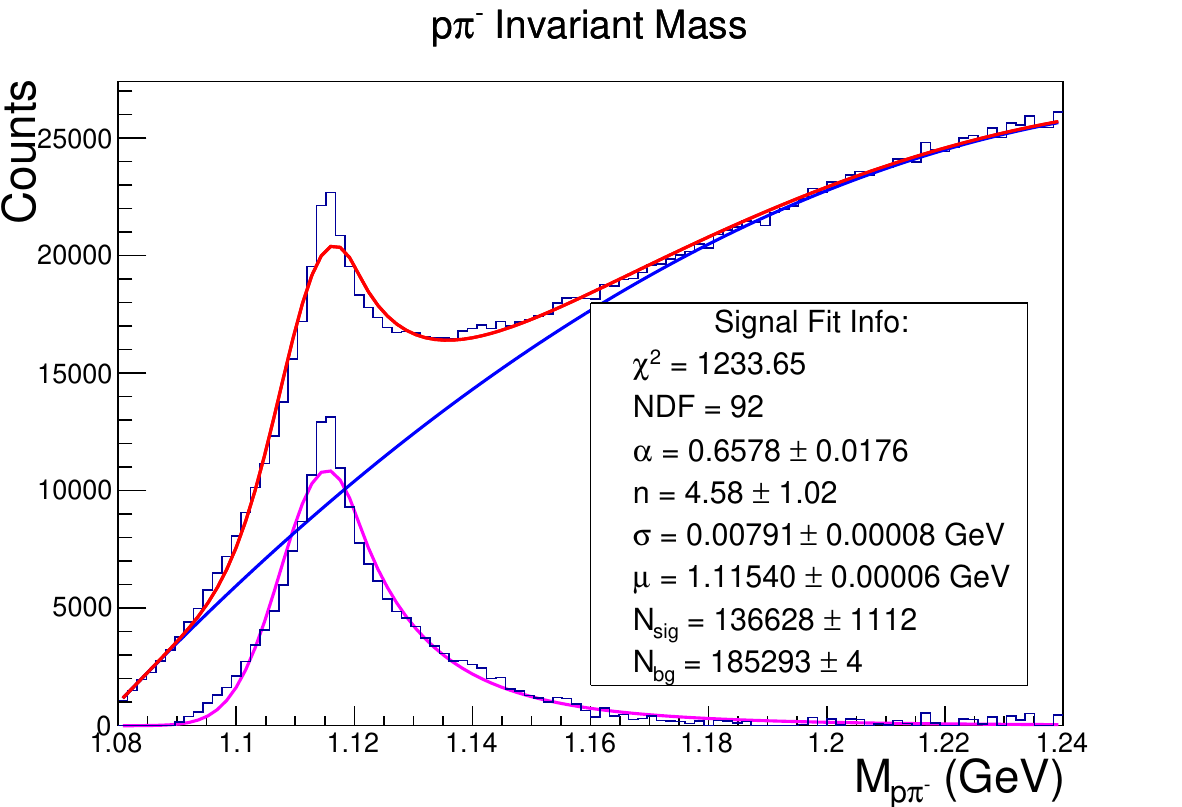}
    \caption{\textbf{Comparison between data  and simulation at CLAS12:} Invariant mass spectrums for the $\Lambda^0$ for data (left) and MC (right). Notice the distinct differences in the shapes of background distributions. Domain adaptation attempts to overcome this via training the GNN with an adversarial loss between the two data formats. Figure taken from Ref.~\cite{McEneaney:2023vwp}. Original figure available under 
    \href{https://creativecommons.org/licenses/by/4.0/legalcode}{Creative Commons Attribution 4.0 International}.
    }
    \label{fig:domain}
\end{figure}

In our discussion on domain adaptation, we examined the usage of Graph Isomorphism Networks (GIN)—an implementation of GNNs that maintains injective functions—for $\Lambda$-event tagging at CLAS12. Leveraging adversarial adaptation, this approach effectively addresses the disparities between training (with simulated data) and deployment (using real data) \cite{McEneaney_2022,McEneaney:2023vwp}. 
When designing ML models, it is often convenient to train and/or test models utilizing simulated data. Simulated data provides high-purity samples in which it is possible to correctly tag each detector candidate given ground truth information. However, when the model is deployed on actual detector response variables, it is assumed that the two data schemes are exact matches, and thus a bias can be introduced. Fig.\ref{fig:domain} shows an example in which the target domain (data) does not match the source domain (MC) for the invariant $\Lambda^0$ mass spectrum. 

\begin{figure}[h!]
    \centering
    \includegraphics[width=0.23\textwidth]{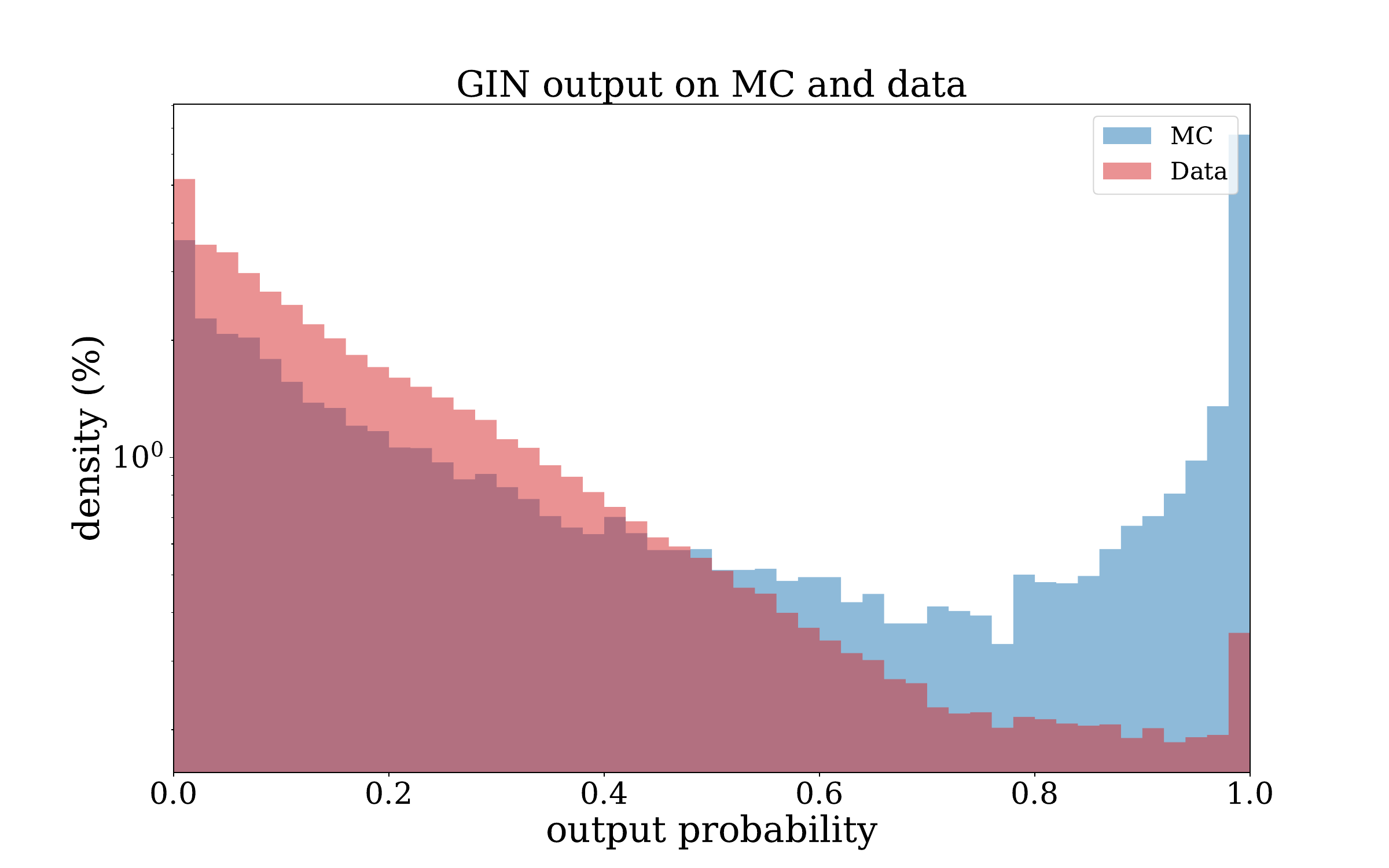}
    \includegraphics[width=0.23\textwidth]{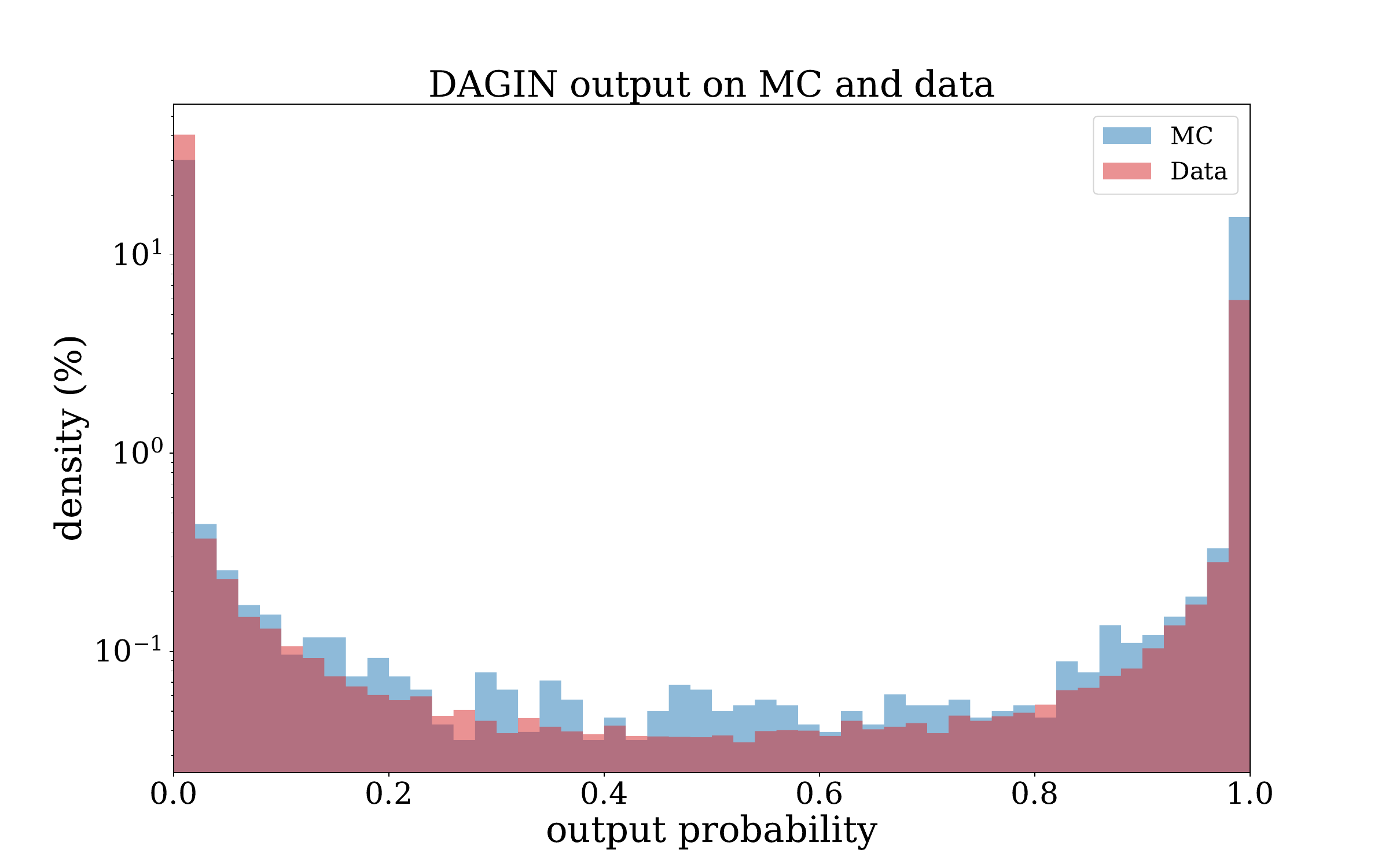}
    \caption{\textbf{Comparison between regular GIN and GIN with domain adaptation (DAGIN) for data and simulation at CLAS12:} The output of the regular GIN (left) shows significant differences between data (blue) and MC (orange). In comparison, the distribution of the outputs for the DAGIN (right) come similar for data and simulation with a Kolmogorov-Smirnov distance 
     for the GIN.
    Figure taken from Ref.~\cite{McEneaney:2023vwp}. Original figure available under \href{https://creativecommons.org/licenses/by/4.0/legalcode}{Creative Commons Attribution 4.0 International}.}
    \label{fig:domainOutput}
\end{figure}

To overcome this, domain adaptation via the utilization of an adversarial training technique was deployed. Information learned from the MC samples should not be disregarded but rather adjusted given the transition to data. After training with an adversarial network with the goal to distinguish between data and simulation, the output distribution of the network becomes significantly more similar as shown in Fig.~\ref{fig:domainOutput}. More details can be found in Ref.~\cite{McEneaney:2023vwp}.

The workshop also spotlighted data-driven inspired approaches \cite{Giroux_2022}. In \cite{Fanelli_2022_FM}, the authors presented a strategy called `Flux+Mutability', which is based on a combination of a conditional autoencoder (cAE), a cMAF, and Hierarchical Density-Based Spatial Clustering of Applications with Noise (HDBSCAN) for one-class classification and anomaly detection. This method has been employed for both $\gamma$/n shower classification in the \gluex barrel calorimeter and detection of potential beyond Standard Model (BSM) di-jet signatures at LHC. 
The F+M algorithm, trained using a single reference class, leverages cAE to filter anomalous events, providing reconstructed features and residuals. The cMAF, fed with these features, generates data for forming a reference cluster, facilitating object-by-object fitting relative to the reference cluster via HDBSCAN. Objects are then labeled using a quantile cut, ensuring class-agnosticity.

\section{Infrastructure and Frontiers}\label{sec:infrastructure}

Artificial intelligence use cases are one of the primary drivers for developing or utilizing new computing infrastructure for the EIC. For example, many scientific domains have developed the foundation for including high performance computing and next generation architectures into their workflows. Similarly, efforts are being made to push ML models closer to the edge of experiments, such as with FPGAs~\cite{Carini:2022vut,Duarte:2019fta,miniskar2022ultra}. The utilization of such hardware requires networks with low computational overhead, in terms of both memory and required floating point operations. Research and development is ongoing to integrate these, and other, new infrastructures into EIC workflows.

One of the biggest challenges currently facing the EIC is the design and development of future-proof infrastructure, viable both currently and in the next decade when data collection commences. Furthermore, any infrastructure developed should also be modular enough to change during the lifetime of the experiments at the EIC, which are expected to be several decades. Designing and deploying a modular computing infrastructure is therefore essential. Defining interfaces between data processing stages such that when new technologies become available, pieces of the overall infrastructure can be updated without disrupting the entire workflow. Lessons can be learned from the LHC, whose computing infrastructure was designed nearly two decades before the accelerator facility began operation. Frameworks did not necessarily just stop working at the LHC as the facility moved farther into the operations phase but rather became inefficient at using the available resources as those resources changed over time which then necessitated changes in the overall infrastructure~\cite{Rohr:2022xcs}. With a sufficiently modular design, pieces that become inefficient could be replaced by new efforts that, for example, take advantage of GPU architectures that were not envisioned to play a large role at the time of design. 

It is also important to consider the role of technologies that are in the early stages of application towards High Energy Physics (HEP) and Nuclear Physics (NP) workflows, such as quantum computing. Continuously checking in and scheduling reviews of the state of technologies, for example when some milestone has been reached, will help assess how applicable they are for workflows at the EIC. Scheduling these ``check ins'' regularly, and starting them early, will help prepare for their possible integration. As an example, a few decades ago GPUs were not expected to be as computationally valuable as they are currently; therefore, it is essential to  remain proactive in evaluating emerging technologies given the timescale of the EIC.

Often times, when designing computing infrastructure, only the hardware and associated software are considered during framework development. However, it is also important to consider the workforce, specifically, how to develop and retain the people necessary to successfully design and implement a computing infrastructure that will serve the EIC science program for its entirety. Building a diverse and interdisciplinary team will help bring technical expertise from computing and physics domains necessary for hardware, algorithm, and physics development. The EIC is a facility that is poised to develop such collaborations due to the size of the project and the necessary cross-cutting challenges that must be overcome for its success, especially with regard to the implementation of ML algorithms in data analysis workflows. Large collaborations, such as those at the EIC, can provide a platform for approaching difficult computational problems; as an example, the Worldwide LHC Computing Grid was created to address the challenges of data collection and processing at the LHC~\cite{Shiers:2007eye}. To develop an interdisciplinary team, connections need to be forged, commonly generated through conferences and workshops. At forums such as these, scientists from a variety of domains are able to discuss approaches to the same problem from the different perspectives their expertise offers. Developing a computing infrastructure that can serve the EIC must include hardware, software, and an interdisciplinary team that is capable of designing, implementing, and maintaining the infrastructure needed to serve the lifetime of the EIC project.

The emergence of Generative Pretrained Transformers (GPT) has offered new potential within the realm of AI for the EIC. With its capability to understand and generate human-like text based on the context provided, GPT models can be pivotal in data interpretation, document generation, and even hypothesis formulation for EIC science and NP at large. This deep learning-based model can sift through large amounts of data, detect patterns, and identify key insights faster and more efficiently than traditional methods, driving further advancements in the field.
At the AI4EIC workshop in October 2022, a month prior to the release of chatGPT, the potential of AI applications in nuclear and particle physics was underlined. The advent of chatGPT further emphasizes this potential, illuminating a promising future where AI tools like GPT can accelerate scientific discovery by automating and enhancing various facets of research in the EIC community.


\section{Streaming Readout}\label{sec:sro}

\begin{figure*}
    \centering
    \includegraphics[width=0.90\textwidth]{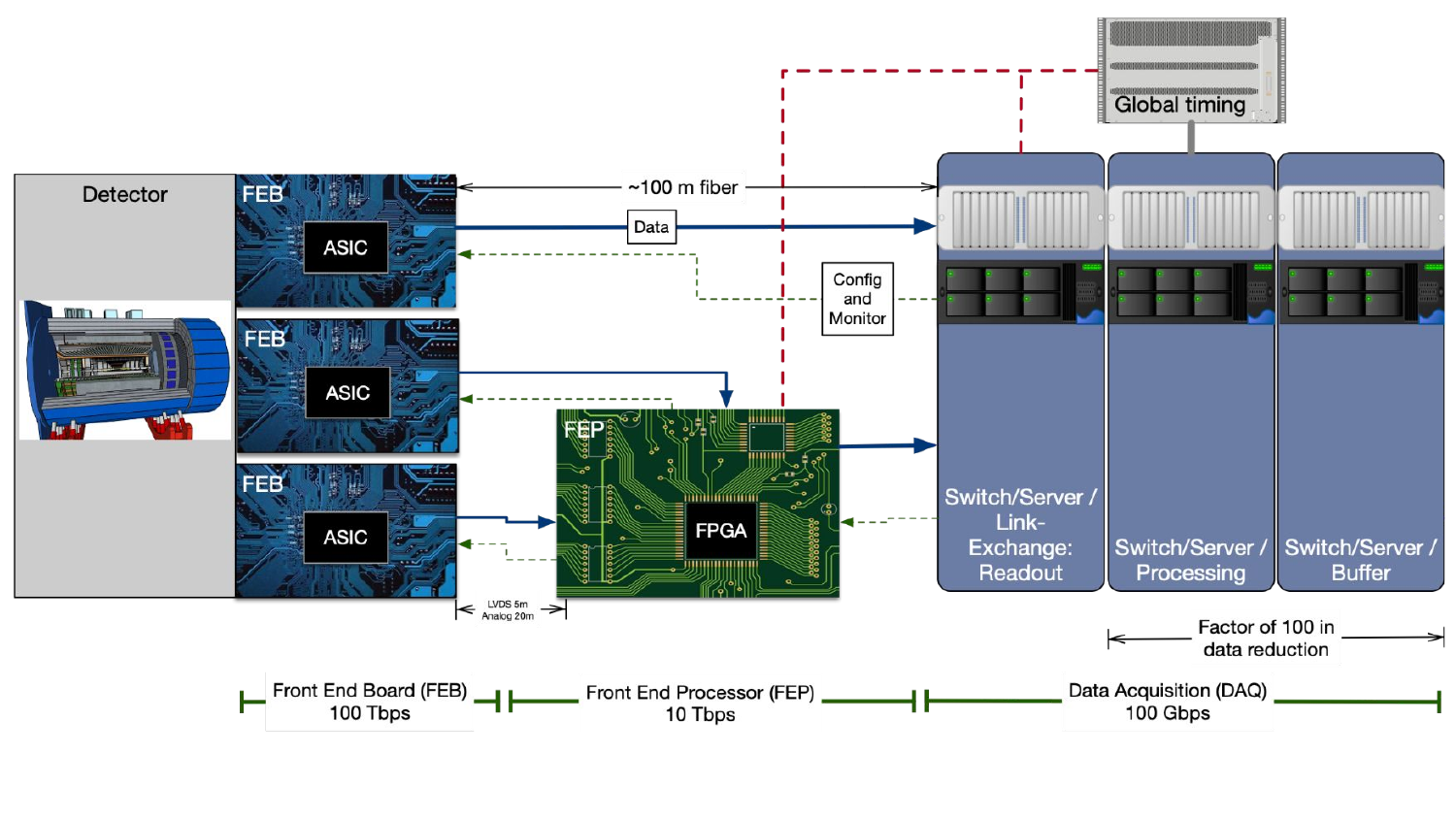}
    \caption{\textbf{Conceptual SRO DAQ System:} The deployment of DAQ electronics is generally segmented by location, comprising the Front End Electronics (FEE) modules adjacent to the detector, the Front End Processor (FEP) boards for digitizing or reformatting detector data, and Stream Aggregator Boards (SAB) located in the hall for bundling streams, with online filtering and monitoring carried out in the counting room. For additional details, refer to \cite{bernauer2023scientific}.}
    \label{fig:SRO-DAQ}
\end{figure*}

SRO is rapidly becoming the go-to paradigm for readout processes in contemporary nuclear and high energy physics experiments. Unlike traditional or pipelined methods that rely on hardware signals for initiating data conversion into the digital realm or marking time regions of interest within close-memory buffers, an SRO data acquisition system incessantly converts and streams detector data to potentially heterogeneous computing systems. The retention of data is determined by software, with possible acceleration by FPGAs or Application-Specific Integrated Circuits (ASICs).
Fig. \ref{fig:SRO-DAQ} provides a conceptual overview of a potential system configuration.

The SRO scheme promptly avails all detector information in digital form, paving the way for AI-powered tagging and filtering algorithms to be employed early in the data collection stage. By making raw data accessible to high-level reconstruction frameworks—typically written in languages such as C++, Python, or Java—SRO allows for the utilization of standard AI tools without necessitating bespoke adjustments for dedicated hardware.
A wide array of system scales and implementations exist, ranging from systems that  record all data to disk, to those that conduct high-level analyses while data is in transit, only preserving high-level physics objects.

SRO has already been implemented in numerous experiments at the LHC (see, \textit{e.g.}, \cite{aaij2020allen,perez201640,mitra2019trigger,migliorini2023trigger}) and has been officially designated as the chosen paradigm for the EIC, as evidenced in the EIC Yellow Report \cite{khalek2021science}. In the case of the sPHENIX experiment, the conventional triggered readout system is augmented with a streaming system for principal detectors, a strategy that permits the exploration of physics phenomena that would otherwise be missed by a triggered system. Similarly, various experiments at Jefferson Lab are experimenting with partial SRO solutions, thereby paving the way towards a comprehensive transition to a full SRO design \cite{ameli2022streaming,furletov2022machine,barbosa2023development}.

The flexible data routing in a streaming readout system enables new or eases the implementation of various quality control and time-to-paper improvements. For example, the INDRA-ASTRA lab at Jefferson Lab is developing techniques to move analysis tasks into the readout \cite{Diefenthaler_2022}. ML has a role here, especially in automatic anomaly detection, for example by using the Adaptive Windowing (ADWIN) technique \cite{bifet2007learning}. In general, streaming readout blurs the lines between online and offline analysis, with the goal to fuse these together as much as possible. 

Similarly, ML has been used for online calibration of the \gluex Central Drift Chamber (CDC) monitoring gain and time-to-distance conversion factors \cite{Britton_2022, jeske2022ai}. 
Implementation of real-time (or quasi-) detector calibration is an essential component of SRO supremacy with-respect-to conventional triggered Data Acquisition systems (DAQs). 
HDBSCAN, a form of unsupervised hierarchical clustering detailed in Section \ref{sec:reco_pid}, has been employed for clustering non-calibrated data from the CLAS12 forward tagger calorimeter in SRO mode \cite{Bondi_2022} and reconstruct the electro-production of $\pi^{0}(\gamma\gamma)$.
To take full advantage of the full off-line data reconstruction framework during data acquisition, raw data need to be calibrated and continuously monitored in order to provide reliable information to tagging/filtering algorithms.
This request represents a great opportunity for AI-supported calibration and monitoring algorithms like those discussed in \cite{Britton_2022}, where the AI system prototype deployed to control and calibrate the \gluex CDC provided good results, paving the way towards a self-calibrating detector.

Machine Learning, particularly GNNs as outlined in Section \ref{sec:reco_pid}, is adept at managing hit and track identification, as showcased in \cite{Furletov_2022}. This study also examines the use of Recurrent Neural Networks (RNNs) and Long Short-Term Memory networks (LSTMs) for track fitting tasks.
Furthermore, ML proves highly effective for noise and background suppression, decreasing the data volume that needs to be transmitted via the readout network and stored on disk. Such implementation yields the most significant impact when deployed early in the readout chain. Considering throughput requirements, there's a clear incentive to incorporate NNs on FPGAs. While packages like hls4ml \cite{fastml_hls4ml} can aid in the implementation, it's noteworthy that not all network topologies are currently supported.

Interesting design advancements have been demonstrated through the development of `bicephalous autoencoders', which offer a lossy compression scheme that retains critical information while suppressing noise, as illustrated in \cite{Huang_2022, huang2021efficient}. 
Current studies are also exploring the use of GNNs for heavy-flavor tagging and their implementation on FPGAs within the context of the sPHENIX project \cite{Dean_2022,Dantong_2021}.
The development of a real-time ML FPGA filter for particle identification and tracking in SRO is outlined in \cite{barbosa2023development}.
Generally, it is expected that future FPGA devices will include more IP cores aimed at acceleration of NNs, for example by integration of matrix multiplication capabilities or higher number of DSP slices. However, the field will have to watch the developments closely. Our needs are not the driver for these developments, and it is unclear if the addition of these abilities will go hand in hand with a reduction in uncommitted resources required for data acquisition IP. 
This potentially presents a challenge, as these newly developed NN accelerator cores may not be compatible with the specific data types required for our unique implementations.

ML also drives developments of new compute models like in-memory computing, with low latencies and very good energy consumption. 
It is clear that the field needs to further approaches, techniques and packages to ease the implementation of NN on multiple FPGA architectures over multiple generations and capabilities, and also to ease the transition from a Central Processing Unit (CPU)/GPU implementation onto an FPGA. This must include also verification tools. 
For a streaming readout system, orchestrating a considerable number of nodes is typically required. This circumstance introduces intricate challenges pertaining to system bring-up and configuration, thus necessitating the standardization of communication protocols. One such framework addressing these challenges is APEIRON \cite{Ammendola_2022,ammendola2023apeiron}.

As already mentioned, beside the world effort driven by CERN experiments,  a significant effort is undergoing at Berkeley Nation Laboratory (BNL) and Jefferson Lab (JLab) to test components and concepts of  a suitable SRO DAQ for EIC. Prototypes of a full DAQ SRO chain have been deployed and tested in both controlled (lab) and realistic (on-beam) conditions (see, \textit{e.g.}, \cite{Bondi_2022}). Results are generally positive and even if current SRO schemes are not expected to be final, the experience gained by the EIC community is valuable for understanding limitations, requirements and opportunities of SRO at EIC.

\section{Community efforts}\label{sec:community}

As we progress further into the 21st century, AI stands as a significant driving force of our economy. In the next decade, as the EIC reaches its operational phase, the impact of AI will be more pronounced than ever. It is in this context that we pivot to the concerted community efforts being made to integrate AI into the EIC landscape.

Recognizing the transformative potential of AI, we have initiated a range of educational activities aimed at enhancing its understanding within the EIC community. These activities are intended to not only increase awareness, but also foster a culture of innovation and exploration centered around AI.
One of our core community initiatives includes organizing hackathons designed around specific challenges pertinent to EIC. These hackathons serve as creative platforms for identifying and discussing promising strategies, architectures, and algorithms. By doing so, they present a unique opportunity to unearth solutions that could significantly bolster the EIC physics program.

In the following, we delve into the nuances of these community efforts and elucidate how they are instrumental in shaping the role of AI within the EIC.

\subsection*{Tutorials}

The workshop incorporated a robust outreach and educational aspect, featuring a series of tutorials presented by esteemed AI and machine learning experts drawn from national laboratories, universities, and industry. Furthermore, a hackathon satellite event was organized, adding a practical element to the last day of the workshop. Four comprehensive tutorials were offered, each designed to impart knowledge on key topics in AI and machine learning. The subjects of these tutorials included Multi-objective optimization with BoTorch/Ax, a technique of unfolding, the concept and applications of Graph Neural Networks, and the Machine Learning lifecycle. This educational component, by bridging the gap between theory and practice, played an essential role in enhancing the attendees' understanding and proficiency in these complex domains.

\paragraph{Multi-objective optimization with BoTorch/Ax:}

Solving multi-objective optimization problems in a sample-efficient fashion is key for in particle accelerator design (and far beyond). BoTorch~\cite{design:botorch} is a modular and highly customizable library for Bayesian Optimization with state-of-the-art algorithmic capabilities. Ax~\cite{design:ax} exposes BoTorch's algorithms through a user-friendly interface and provides additional high-level management, storage, and orchestration capabilities.

In this tutorial, we go over some basic hands-on examples of how to use Ax to perform multi-objective Bayesian Optimization via Ax's Service API (an ask/tell interface) on a synthetic problem. This setup is straightforwardly adapted to any actual multi-objective black-box optimization problem with costly evaluations. The full tutorial is available here: \cite{mobotutorial} (slides), \cite{mobotutorialcolab} (colab notebook).

\paragraph{Unfolding:}
Unfolding aims to correct measured observables for detector distortions and provide easy access to theoretical quantities for the broader nuclear and high energy physics community.  Existing unfolding methods  require the usage of histograms and are limited to low-dimensional inputs and outputs. Machine learning can naturally incorporate high-dimensional data to estimate the detector response, providing a more accurate estimation of the measured observable. In this tutorial, we will introduce OmniFold \cite{andreassen2020omnifold}, a machine learning-based method that simultaneously determines the unfolded response of multiple distributions. We present recent results of the application of OmniFold to particle collisions collected by the H1 Collaboration and provide hands-on tutorials on a toy example using normal distributions as well as an example motivated by the EIC, unfolding the kinematics of leptons and hadrons in deep inelastic scattering (DIS).
The Colab notebook is available here \cite{toralesacostaomnifold}. 

\paragraph{Machine Learning Lifecycle:}
The phases of the machine learning life cycle may be thought of as 1) data analysis, 2) experimentation, 3) model reproducibility, 4) deployment, and 5) production monitoring. This tutorial introduces the Machine Learning Operations (MLOps) open-source platform MLFlow \cite{zaharia2018accelerating} and describes MLFlow’s four components to support the ML lifecycle with a specific focus on the MLFlow Tracking component, which is used to record and compare machine learning trials. The python library HyperOpt is also introduced, a library for hyperparameter optimization.
The tutorial is contained within a Colab notebook and uses publicly available data from the 2021 Jefferson Lab hackathon when an imaginary calorimeter, with a single shower and no noise, was simulated.  The problem is easily solvable with a simple neural network and is used only to illustrate the ease of implementing hyperparameter optimization and MLFlow tracking. First, users are guided through a grid search for a “best model” by creating different types of neural networks with different hyperparameters and tracking the results. Then comparing the results to determine the best-performing model.

After users understand the few lines of code needed to implement the grid search implementation of MLFlow Tracking they are introduced to the HyperOpt python library, the concept of the “search space”, the “minimization function”, and how to combine these concepts into the model’s training function and how to track the best performing hyperparameters using MLFlow.

The Colab notebook is available \cite{mlopstutorial}.

\paragraph{Graph Neural Network:}
Many real-world data, such as social networks, molecules, roadway maps, cellular biological pathways and so on, are sparse.
It is more effective to represent such data as a graph representing relationship among entities.
Graph Neural Networks feature permutation invariance on handling graph data. 
Similar to translation invariance in convolutional neural networks, where the kernel remains the same at different locations of an image,
graph neural network is invariant to how the nodes are ordered.

This tutorial is a self-contained Colab notebook that goes through a basic graph neural network on solving a regression problem: determine the solubility given a molecule structure.
In particular, the tutorial dives deep in practical GNN techniques such as how to generate node features, how to construct a graph convolution layer, how to batch multiple graphs in a mini-batch and so on.
At the end, users can pick different hyper-parameters to train and evaluate the GNN model.
The Colab note is available here \cite{gnntutorial}.


\subsection*{Hackathon}



The format of the hackathon was hybrid and international (both local and remote participation), with more than 30 participants connected from around the world (America, Asia and Europe, mainly) grouped in 10 different teams competing to solve the assigned problems.  
Access to cloud computing resources has been provided during the event, and each team was endowed with an Amazon Web Services (AWS) g5.12xlarge instance, 4 Nvidia a10g GPUs, 48 vCPUs, 192GB Ram, 3.9 TB of disk.

For this hackathon we proposed problems with increased level of difficulty and that are deemed to be solvable in a one-day event, starting from a problem that is accessible to everyone.
We focused on the dual-radiator Ring Imaging Cherenkov detector under development as part of the particle-identification system at the future \epic detector at EIC. 
Data have been produced using the \epic software stack.

\begin{figure}[!]
    \centering
    \includegraphics[scale = 0.25]{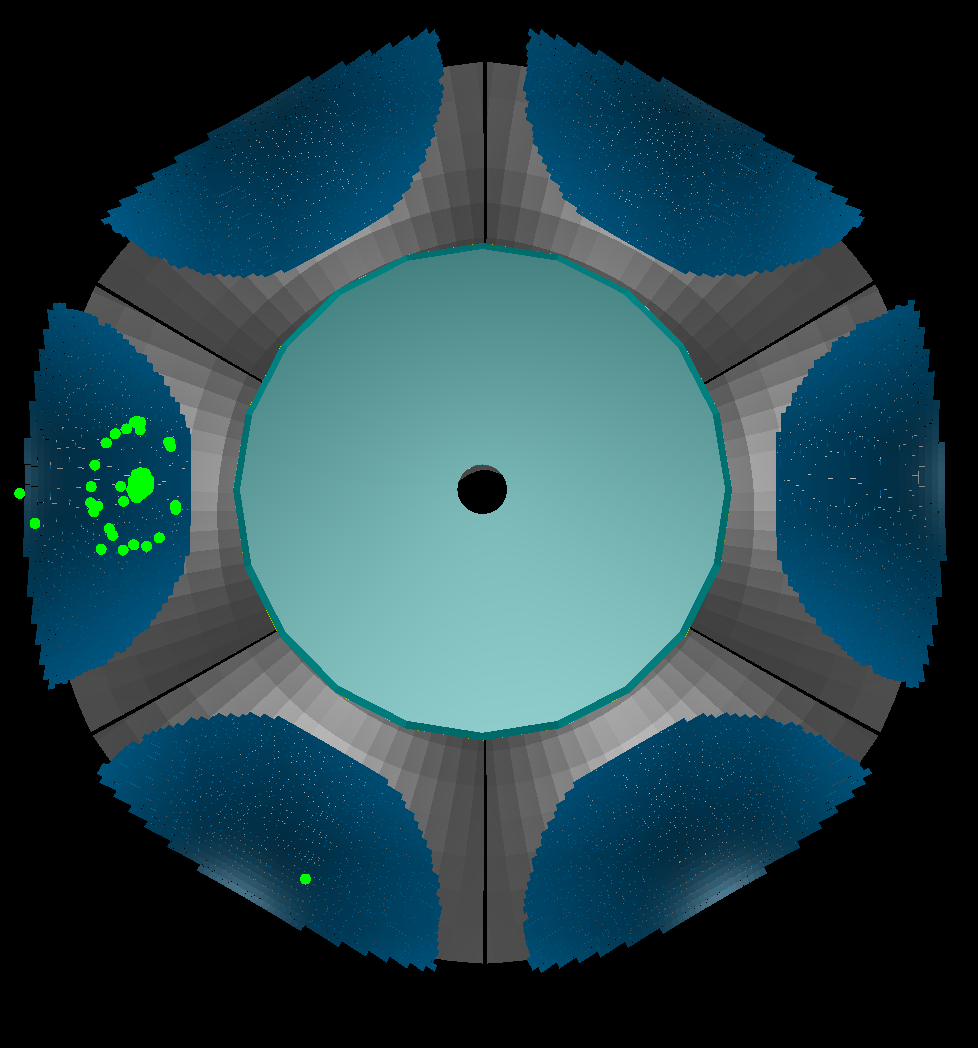 }
    \caption{A sample dRICH $\pi^{+}$ event visualized using \epic framework.}
    \label{fig:dRICHGeom}
\end{figure}

The hackathon was structured around three problems, each escalating in complexity. Initially, we selected a momentum range around 15 GeV as our foundation problem. This range is significant as it corresponds to a momentum zone where both aerogel and gas radiators can potentially contribute to the $\pi/K$ separation. In order to raise the level of challenge, we embedded realistic photon yields. An exemplar $\pi^{+}$ event as detected in dRICH is depicted in Fig.~\ref{fig:dRICHGeom}.
Moving to the second problem, we expanded the scope by varying the momentum range and altering the positions of the pions and kaons within the dRICH.
For the ultimate challenge, the final problem introduced a layer of complexity with a set of random noise hits, making it the most demanding among all three problems.
Documentation and data sets have been made available on Zenodo \cite{Fanelli_2022}.
Despite the inherent `simplicity' of the problems, given the approximations made as explained in this document, this event can potentially become a first step towards machine learning/deep learning application for PID with the dRICH.
At the end of the hackathon event, the best solutions provided were all machine learning/deep learning-based, they were quite original, and they outperformed other solutions based on more `classical' approaches (like cut-based analyses). 
Though an initial foray into leveraging machine learning and deep learning for PID with the dual-RICH, these studies unequivocally point towards the potential these novel approaches hold for reconstruction and PID within the \epic dual-RICH framework. This endeavor proved to be a valuable learning opportunity, especially for students, and intriguingly, it showcased the potential edge that contemporary AI/ML methods hold over traditional strategies for PID in imaging Cherenkov detectors, as discussed in \cite{fanelli2020machine} and references therein.


\section{Conclusions}\label{sec:conclusions}

The AI4EIC workshop has successfully highlighted the critical role that AI/ML play in the design and execution of the Electron Ion Collider (EIC). The event was organized into multiple sections, each focusing on various aspects of the EIC science and the connections with AI/ML applications, providing participants with a comprehensive understanding of these complex topics.

As a community, we are enthused to see the funding opportunities for AI/ML in relation to the EIC, which is promising to yield significant results in the coming years. This financial commitment will undoubtedly contribute to the acceleration of research and innovation within the field.
AI4EIC is not only a platform for showcasing the remarkable advancements and progress in AI, but also plays a vital role in increasing AI literacy. By disseminating the knowledge and understanding of AI across the community, we hope to inspire more individuals and institutions to engage with this technology.
We are also pleased to collaborate with the ePIC experiment at the EIC. This partnership is set to bring new perspectives and opportunities for progress, strengthening the role of AI in our initiatives.
The success of our educational activities, such as hackathons and schools, affirms the effectiveness of these strategies. We are committed to continuing such events, facilitating an environment that encourages learning, innovation, and collaboration.

We are excited to host the third AI4EIC workshop at The Catholic University of America (CUA) in Washington D.C., in November 2023. We look forward to building on the success of this workshop, furthering the discussion, and driving the integration of AI into the EIC's future endeavors.
As plans for future events, we anticipate formats such as conference, schools, and data challenges.

\bmhead{Acknowledgments}

The authors would like to acknowledge AWS for providing the cloud computing resources for the hackathon event. We also, thank the College of William and Mary for their support to the hackathon and for sponsoring the prizes.
%

\bibliography{sn-bibliography}


\end{document}